\documentclass{jpp}
\usepackage{graphicx}
\usepackage{epstopdf, epsfig}
\usepackage{hyperref}
\usepackage{color}

\newcommand{\prl}{    {Phys. Rev. Lett.}}

\newcommand{\apj}{    {Astrophys. J.}}
\newcommand{\apjl}{    {Astrophys. J. Lett.}}

\newcommand{\ssr}{    {Space Sci. Rev.}}
\newcommand{\planss}{    {Plan. Sp. Sci.}}
\newcommand{\jgr}{    { J. Geophys. Res.}}
\newcommand{\grl}{    { Geophys. Res. Lett.}}
\newcommand{\solphys}{    { Solar Physics}}
\newcommand{\nat}{    {Nature}}
\newcommand{\pre}{    {Phys. Rev. E}}

\shorttitle{From Hamiltonian resonance theory to phase space mapping}
\shortauthor{A. V. Artemyev, et al.}

\title{Long-term dynamics driven by resonant wave-particle interactions: from Hamiltonian resonance theory to phase space mapping}
\author{Anton V. Artemyev\aff{1,2}
  \corresp{\email{aartemyev@igpp.ucla.edu}},
  Anatoly I. Neishtadt\aff{3,2},
 Alexei. A.  Vasiliev\aff{2},
  Xiao-Jia  Zhang\aff{1},
  Didier Mourenas\aff{4} \and
  Dmitri Vainchtein\aff{5,2}}
\affiliation{
\aff{1}Institute of Geophysics and Planetary Physics, UCLA, Los Angeles, California 90095, USA;
\aff{2}Space Research Institute of the Russian Academy of Sciences (IKI), 84/32 Profsoyuznaya Str., Moscow 117997, Russia;
\aff{3}Department of Mathematical Sciences, Loughborough University, Loughborough LE11 3TU, United Kingdom;
\aff{4}Laboratoire Mati\`ere sous Conditions Extr\^emes, Paris-Saclay University, CEA, Bruy\`eres-le-Ch\^atel, France;
\aff{5}Nyheim Plasma Institute, Drexel University, Camden, NJ, USA;
}
\begin{document}

\maketitle

\begin{abstract}
In this study we consider the Hamiltonian approach for the construction of a map for a system with nonlinear resonant interaction, including phase trapping and phase bunching effects. We derive basic equations for a single resonant trajectory analysis and then generalize them into the map in the energy/pitch-angle space. The main advances of this approach are the possibility to consider effects of many resonances and  to simulate the evolution of the resonant particle ensemble on long time ranges. For illustrative purposes we consider the system with resonant relativistic electrons and field-aligned whistler-mode waves. The simulation results show that the electron phase space density within the resonant region is flattened with reduction of gradients. This evolution is much faster than the predictions of quasi-linear theory. We discuss further applications of the proposed approach and possible ways for its generalization.
\end{abstract}

\section{Introduction}
The resonant wave-particle interaction is known to be one of the main drivers of dynamics of such space plasma systems as planetary radiation belts \citep[e.g.,][]{Thorne10:GRL, Menietti12}, collisionless shock waves \citep[e.g.,][]{Balikhin97, Wilson07, Wilson12, Wang20:apjl}, auroral acceleration region \citep[e.g.,][]{Chaston08, Watt&Rankin09:prl, Mauk17:nature}, and solar wind \citep[e.g.,][]{Krafft&Volokitin16, Kuzichev19, Tong19:ApJ, Yoon19, RobergClark19}. The classical quasi-linear theory \citep{Vedenov62, Drummond&Pines62} and its generalizations for inhomogeneous plasma systems \citep{Ryutov69, bookLyons&Williams} describe well charged particle resonant interaction with low-amplitude broadband waves \citep{Karpman74:ssr, Shapiro&Sagdeev97, Tao12, Camporeale&Zimbardo15, Allanson20}.

One of the important examples of application of the quasi-linear theory is the Earth radiation belt models that describe energetic electron acceleration and losses due to resonances with electromagnetic whistler-mode waves and electromagnetic ion cyclotron (EMIC) waves \citep[see reviews][and references therein]{Thorne10:Nature, Shprits08:JASTP_local, Ni16:ssr, Nishimura10:Science, Millan&Thorne07}. Moreover, the natural inhomogeneity of the background magnetic field and plasma density in the radiation belts can significantly weaken the conditions of applicability of the quasi-linear theory \citep{Solovev&Shkliar86, Albert01, Albert10}. However, this theory meets difficulties in describing resonances with sufficiently intense waves \citep{Shapiro&Sagdeev97}, when the nonlinear effects of phase trapping and phase bunching become important \citep{Omura91:review, Shklyar09:review, Albert13:AGU, Artemyev18:cnsns}. Indeed, sufficiently intense whistler-mode waves are frequently observed in the radiation belts \citep{Cattell08, Cattell11:Wilson, Agapitov14:jgr:acceleration} and contribute significantly to wave statistics \citep{Zhang18:jgr:intensewaves, Zhang19:grl, Tyler19}. Theoretically, phase trapping and bunching (also called nonlinear scattering) effects are responsible for fast acceleration \citep[e.g.,][]{Demekhov06, Demekhov09, Omura07, Hsieh&Omura17, Hsieh20} and losses \citep[e.g.,][]{Kubota15,Kubota&Omura17, Grach&Demekhov20} of energetic electrons and for the generation of coherent whistler-mode waves \citep{Demekhov11, Katoh14, Katoh&Omura16, Tao14:code, Omura08, Omura13:AGU, Nunn&Omura12}. There are many observational evidences of such nonlinear resonant wave generation \citep{Titova03, Cully11, Tao12:sweeprate, Mourenas15} and of the related electron acceleration/losses \citep[e.g.,][]{FosterGRL14, Agapitov15:grl:acceleration, Mourenas16, Chen20:microbursts}. 

The quasi-linear diffusion theory describes a sufficiently weak scattering in energy/pitch-angle space and operates with a Fokker-Planck diffusion equation for the charged particle distribution function \citep{Andronov&Trakhtengerts64, Kennel&Engelmann66, Lerche68}. In contrast to this description, the nonlinear phase trapping assumes a fast transport in energy/pitch-angle space \citep[e.g.,][]{Artemyev14:grl:fast_transport, Furuya08}, when even a single resonant interaction changes significantly the electron's energy/pitch-angle \citep[e.g.,][]{Albert13:AGU, Artemyev18:cnsns}. This essentially non-diffusive process cannot be directly included into the Fokker-Planck equation. One possible approach is the construction of an operator that would describe fast charged particle jumps in the energy/pitch-angle; this operator can be constructed with the numerical (test-particle) approach \citep[e.g.,][]{Hsieh&Omura17:radio_science, Zheng19:emic} or with the analytical calculation of jumps' probabilities \citep[e.g.,][]{Vainchtein18:jgr}. The main advantage of this approach is the inclusion of almost arbitrary (as realistic as needed) wave spectrum and characteristics \citep[e.g., wave modulation and frequency drifts, see][]{Kubota&Omura18, Artemyev19:cnsns, Hiraga&Omura20}. The main disadvantages are an accumulation of numerical errors with running time, and the almost intractable fine details of the energy/pitch-angle space binning needed to simultaneously resolve large jumps due to trapping and small changes due to drift/diffusion. 

An alternative approach to the construction of such an operator is a generalization of the Fokker-Planck equation to include effects of phase trapping and phase bunching  \citep{Solovev&Shkliar86, Artemyev16:pop:letter, Artemyev17:pre}. This approach is based on a fine balance of trappings and bunchings for a single wave system \citep[e.g.,][]{Shklyar11:angeo, Artemyev19:pd}. The main advantage of this approach is that the evolution of charged particle distribution function can be investigated in arbitrary details in presence of phase trapping, phase bunching, and diffusion \citep[e.g.,]{Artemyev18:jpp, Artemyev19:pd}. The main disadvantage is that there is no straightforward generalization of this approach for multi-wave (multi-resonance) systems. A single-wave resonance results in charged particle transport in the energy/pitch-angle space along 1D curves, so-called resonance surface curves \citep[e.g.,][]{bookLyons&Williams, Summers98}, and the Fokker-Planck equation with trapping was derived for such a quasi-1D system \citep{Artemyev16:pop:letter}. 

Another alternative for the description of charged particle distribution evolution driven by nonlinear wave-particle interaction (phase trapping and bunching) is the mapping technique that describes the characteristics of the Fokker-Planck equation \citep{book:VanKampen03}. The classical example of this approach is the Chirikov map \citep{Chirikov79}, which describes particle diffusion and is widely used for systems with wave-particle resonances \citep[e.g.,][]{Vasiliev88, Zaslavskii89:jetp, Benkadda96, Khazanov13, Khazanov14}. Such a map has been constructed for a single-wave system with phase-trapping and phase bunching effects \citep{Artemyev20:pop}. In this study we show the generalization of this map for a multi-resonance system. 

We consider a strong magnetic field system, where charged particle motion is well gyrotropic and magnetic moments are well conserved away from the resonances. Thus, 3D velocity space can be reduced to 2D energy/pitch-angle space. The mapping for this space should describe 2D charged particle motion due to energy/pitch-angle jumps with the time-intervals between jumps equal to the interval between passages through the resonances. Diffusive jumps (with zero mean values) and jumps driven by nonlinear phase bunching and phase trapping depend on the resonant phase $\varphi_R$, i.e. a variable proportional to the particle gyrophase, which changes fast. In low wave intensity systems this phase is randomly distributed over entire ($\varphi_R\in[0,2\pi]$) range, and the phase dependence $\sim\sin\varphi_R$ can be directly included into the map \citep{Vasiliev88, Zaslavskii89:jetp, Benkadda96, Khazanov13, Khazanov14}. The phase bunching and phase trapping operate in certain $\varphi_R$ ranges \citep[e.g.,][]{Albert93, Itin00, Grach&Demekhov20}, whereas jumps depend on $\varphi_R$ quite nonmonotonically (see \citet{Artemyev14:pop, Artemyev18:cnsns}). However, due to phase randomization between two successive resonances (see Appendix in \citep{Artemyev20:pop}), the phase-dependence can be reduced to a simplified determination of ranges corresponding to phase trapping $\varphi_R\in[0,2\pi\Pi]$ and phase bunching $\varphi_R\in[\Pi,2\pi]$ where $\Pi< 1$ is the probability of trapping (see, e.g., \citet{Artemyev18:cnsns}). The phase gain between two resonances is a large value depending on particle energy and pitch-angle, but this dependence can be omitted in the leading approximation (see discussion in \citet{Artemyev20:pop}). Therefore, in this study we consider charged particle transport in the energy/pitch-angle space due to nonlinear resonant interaction under assumption of resonant phase randomization (limitations of this assumption have been studied in \citet{Artemyev20:rcd}).

The paper structure includes a description of the basic system properties and examples of multi-resonant systems observed in the Earth’s radiation belts (Sect. 1). We present three examples: with two whistler-mode waves providing two cyclotron resonances, with one oblique whistler-mode wave providing cyclotron and Landau resonances, and with one whistler-mode wave and one EMIC wave providing two different cyclotron resonances. Then we focus on the first example and construct the map for this system (Sect. 2). Theoretical results derived from this map are verified with test particle simulations. At the end of the paper we discuss  the constructed map and possible extensions of the proposed approach (Sect. 3).

\section{Basic system properties} 
The Hamiltonian of a relativistic charged particle (e.g., an electron with rest mass $m_e$ and charge $-e$) moving in the 2D inhomogeneous magnetic field of the Earth dipole and interacting with electromagnetic waves (in the low amplitude limit with the wave energy $U_w$ much smaller than electron energy $\sim m_ec^2$, where $c$ is the speed of light) can be written as \citep[e.g.,][]{Albert13:AGU, Artemyev18:jpp}:
\begin{eqnarray}
H &=& m_e c^2 \gamma  + U_w \left( {s,I_x } \right)\sin \left( {\phi  \pm n\psi } \right)\nonumber\\ \gamma  &=& \sqrt {1 + \frac{{p_\parallel ^2 }}{{m_e^2 c^2 }} + \frac{{2I_x \Omega _{ce} }}{{m_e c^2 }}} ,
\label{eq01}
\end{eqnarray}
where two pairs of conjugate variables are $(s, p_\parallel)$ (the field-aligned coordinate and momentum) and $(\psi, I_x)$ (gyrophase and momentum $I_x=c\mu/e$; $\mu$ is the classical magnetic moment). The electron gyrofrequency $\Omega_{ce}=eB_0/m_ec$ is determined by the background magnetic field $B_0(s)$, given by, e.g., the reduced dipole model \citep{Bell84}. The sign $\pm$ in front of $\psi$ is determined by the wave polarization: $+$ for whistler-mode waves interacting with electrons and $-$ for EMIC waves interacting with electrons. The resonance number is $n=0,\pm1, \pm2...$. The wave vector ${\bf k}=(k_{\parallel}(\omega,s), k_{\perp}(\omega,s))$ is given by cold plasma dispersion equation \citep{bookStix62} for a constant wave frequency $\omega$ (i.e., $\partial \phi/\partial s=k_\parallel$, $\partial \phi/\partial t=\omega$). For a finite angle $\theta={\rm arctan}( k_\perp/ k_\parallel)$ between the wave vector and the background magnetic field the wave amplitude in Hamiltonian (\ref{eq01}) takes the form \citep{Albert93, Tao&Bortnik10, Nunn&Omura15, Artemyev18:jpp}:
\begin{eqnarray}
 U_w  &=& \sqrt {\frac{{2I_x \Omega _{ce} }}{{m_e c^2 }}} \frac{{eB_w }}{k}\sum\limits_ \pm  {\frac{{\cos \theta  \pm C_1 }}{{2\gamma }}J_{n \pm 1} \left( {\sqrt {\frac{{2I_x k^2 }}{{m_e \Omega _{ce} }}} \sin \theta } \right)}  \nonumber \\
\label{eq02}\\ 
  &+& \frac{{eB_w }}{k}\left( {\frac{{p_\parallel }}{{\gamma m_e c}} + C_2 } \right)J_n \left( {\sqrt {\frac{{2I_x k^2 }}{{m_e \Omega _{ce} }}} \sin \theta } \right)\sin \theta  \nonumber 
 \end{eqnarray}
where $B_w$ is the wave magnetic field amplitude, $C_{1,2}$ are functions of wave dispersion and $\theta$, and $J_n$ are Bessel functions. Equation (\ref{eq02}) shows that for field-aligned waves $\theta=0$ there is only one cyclotron resonance $n=-1$: $U_{w}=\sqrt{2I_x\Omega_{ce}/m_ec^2}eB_w/k\gamma$ (with $C_1=1$ for $\theta=0$, see \citep{ Tao&Bortnik10}). For oblique wave propagation $\theta\ne 0$ the whole set of resonances with different values of $n$ is present.

\subsection{Field-aligned whistler waves}
\begin{figure}
\centerline{\includegraphics[width=1.0\textwidth]{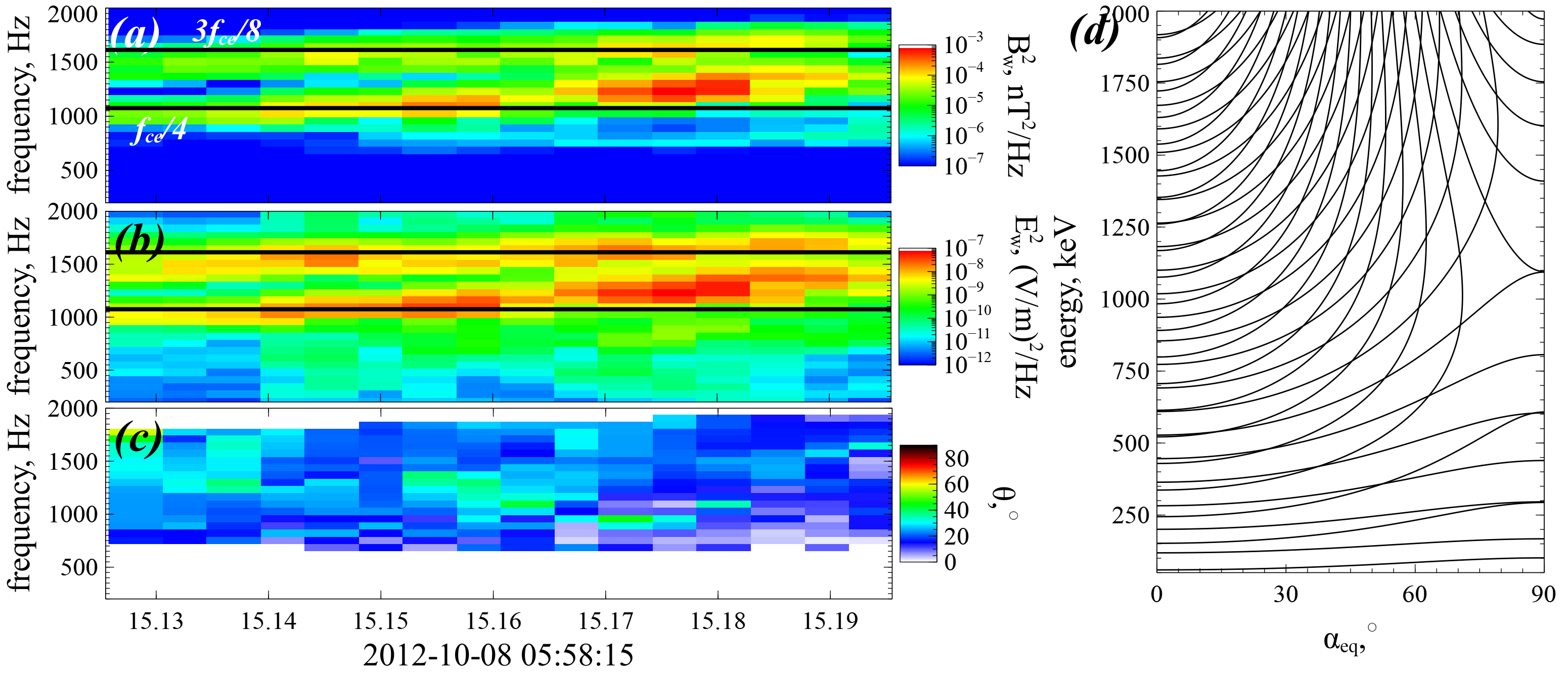}}
  \caption{Observational example of two whistler-mode waves by Van Allen Probe A \cite{Mauk13}: magnetic field spectrum (\textit{a}) and electric field spectrum (\textit{b}) are obtained from EMFISIS measurements \cite{Kletzing13}, wave-normal angle (\textit{c}) is estimated using  the singular value decomposition method \cite{Santolik03}. Resonance curves for the system with two observed whistler-mode waves (\textit{d}).}
\label{fig1}
\end{figure}

Let us start with the system of two field-aligned whistler waves with the Hamiltonian:
\begin{equation}
H = m_e c^2 \gamma  + \sqrt {\frac{{2I_x \Omega _{ce} }}{{m_e c^2 }}} \frac{e}{\gamma }\sum\limits_{i = 0,1}^{} {\frac{{B_{w,i} }}{{k_i }}\sin \left( {\phi _i  + \psi } \right)} 
\label{eq03}
\end{equation}
where $\partial\phi_i/\partial s=k_i=k_i(\omega_i,s)$ with the two different wave frequencies $\omega_i$. Figure \ref{fig1} shows an example of such system observations. THEMIS spacecraft measures waves within the whistler-mode frequency range ($f\in[0.1, 1] f_{ce}$; $f_{ce}=\Omega_{ce}/2\pi$): there are two clear maxima in the magnetic and electric field spectra at $f\sim f_{ce}/4$ and $f\sim 3f_{ce}/8$ (see panels (a) and (b)). Both waves propagate along the background magnetic field: panel (c) shows $\theta$ as a function of the frequency. These double-peak spectra are quite typical for  whistler-mode waves in the inner magnetosphere \citep[see, e.g.,][]{Meredith07, Ma17:vlf, Crabtree17:jgr, Zhang20:grl:frequency, He20, Yu20}.

To study electron energy/pitch-angle variation in the system with Hamiltonian (\ref{eq03}), we follow the standard procedure \citep{Neishtadt06, Neishtadt14:rms} and introduce the wave phases as new canonical variables, $\varphi_i=\phi_i+\psi$, with the generating function:
\begin{equation}
W = sP + \left( {\int {k_0 (\tilde s)d\tilde s}  - \omega _0 t + \psi } \right)I_0  + \left( {\int {k_1 (\tilde s)d\tilde s}  - \omega _1 t + \psi } \right)I_1 
\label{eq04}
\end{equation}
This function gives new variables: $P=p-k_0I_0-k_1I_1$, $S=s$ (we keep $s$ notation), $I_x=I_0+I_1$, and new Hamiltonian $H_I=H+\partial W/\partial t=H-\omega_0I_0-\omega_1I_1$ 
\begin{eqnarray}
 H_I  &=&  - \omega _0 I_0  - \omega _1 I_1  + m_e c^2 \gamma  + \sqrt {\frac{{2\left( {I_0  + I_1 } \right)\Omega _{ce} }}{{m_e c^2 }}} \frac{e}{\gamma }\sum\limits_{i = 0,1}^{} {\frac{{B_{w,i} }}{{k_i }}\sin \varphi _i }  \nonumber \\ 
\label{eq05}\\
 \gamma  &=& \sqrt {1 + \frac{{\left( {P + k_0 I_0  + k_1 I_1 } \right)^2 }}{{m_e^2 c^2 }} + \frac{{2\left( {I_0  + I_1 } \right)\Omega _{ce} }}{{m_e c^2 }}}  \nonumber
 \end{eqnarray}
Hamiltonian $H_I$ describes a conservative system ($H_I={\rm const}$; without loss of generality we take $H_I={0}$) with three degrees of freedom, i.e., with three pairs of  conjugate variables $(s, P)$, $(\varphi_0,I_0)$, $(\varphi_1, I_1)$. The resonance $\dot\varphi_i=0$ conditions give $I_{0}=I_{0R}(s,P,I_1)$, $I_{1}=I_{1R}(s,P,I_0)$ as solutions of equations $\omega_i=m_ec^2\partial\gamma/\partial I_i = 0$. 
Thus, there are two resonant surfaces. If these surfaces cross  (i.e., at the same $s,P$ electron can have simultaneously $I_{0}=I_{0R}$ and $I_1=I_{1R}$),then electrons can simultaneously be in resonance with the two waves \citep{Shklyar&Zimbardo14, Zaslavsky08:double_resonance}. This quite complicated system would require a separate consideration \citep{bookSagdeev88, Lichtenberg&Lieberman83:book}. Hereafter, we focus instead on the simpler case of well-separated resonances, when resonant surfaces do not cross. Equations  $I_{0}=I_{0R}$ and $I_1=I_{1R}$ together with the condition $H_I={0}$ determine two families of curves 
in in $(s,P)$ plane; values $I_{0}$ and $I_{1}$ are parameters of these families.
Thus, on the curve $I_0=I_{0R}(s,P,I_1)$ there is no change of $I_1$, and on the curve $I_1=I_{1R}(s,P,I_0)$ there is no change of $I_0$. With constant $I_1$ (or $I_0$) the $I_0$ (or $I_1$) variation is directly related to the variation of energy: $-\omega_0I_0-\omega_1I_1+m_ec^2\gamma= 0$. Taking into account that $I_0+I_1=I_x=m_ec^2(\gamma^2-1)\sin^2\alpha_{eq}/2$, we can plot resonance curves \citep[e.g.,][]{bookLyons&Williams, Summers98, Mourenas12:JGR:acceleration}, along which $I_{i}$ change, in the energy/pitch-angle space ($m_ec^2(\gamma-1),\alpha_{eq}$)  (note that we use the equatorial pitch-angle $\alpha_{eq}$ defined at the minimum of $B_0(s)$ field, i.e., at the minimum of $\Omega_{ce}(s)$). Figure \ref{fig1}(d) shows these curves $m_ec^2\gamma-\omega_{i}I_{i}={\rm const}$: each curve of $I_0$ change corresponds to a fixed value of $I_1$, and vice versa. Electrons move along these curves with the time-step of the interval between resonances. Note between resonances both $I_{i}$ and $\gamma$ are conserved, and electrons are moving along adiabatic orbits without wave influence, i.e. energy and pitch-angle change only at the resonances.

\begin{figure}
 \centerline{\includegraphics[width=1\textwidth]{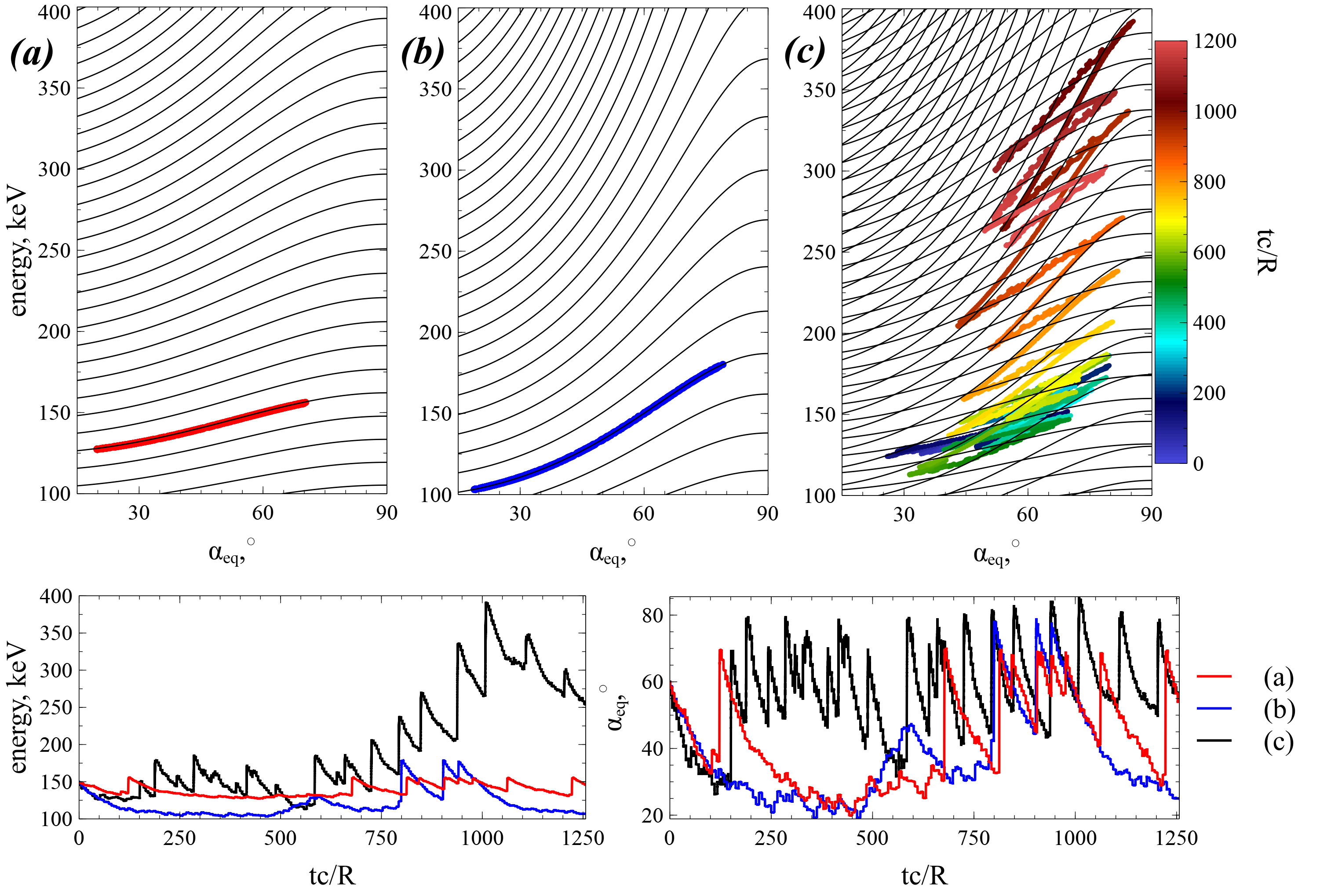}}
  \caption{Top panels show resonance curves (black) and electron trajectories in the energy/pitch-angle space for Hamiltonian (\ref{eq03}): only the first whistler-mode wave (\textit{a}), only the second whistler-mode wave (\textit{b}), both whistler-mode waves  (\textit{c}). In panel (\textit{c}) different colors correspond to different time intervals. The bottom panels show energy and pitch-angle time series for the three trajectories in the top panels.
  We use parameters of dipole field at a radial distance of $\sim 5$ Earth radii. Plasma density is given by the model from \citet{Sheeley01} and constant along magnetic field lines. Wave frequencies are $\omega_0=0.4\Omega_{ce}$, $\omega_1=0.2\Omega_{ce}$. Both wave amplitudes are $300$ pT \citep[see discussion of such wave observations in][]{Zhang18:jgr:intensewaves, Tyler19}. Wave amplitude is distributed along magnetic field line as  $\tanh((\lambda/\delta\lambda_1)^2)\exp(-(\lambda/\delta\lambda_2)^2)$ with $\lambda$ the magnetic latitude ($ds=Rd\lambda\sqrt{1+\sin^2\lambda}\cos\lambda$) and $\delta\lambda_1=2^\circ$, $\delta\lambda_2=20^\circ$. This function fits the observed whistler-mode wave intensity distribution \cite{Agapitov13:jgr}.}
\label{fig2}
\end{figure}

Let us consider electron dynamics in the energy/pitch-angle space for the system with Hamiltonian (\ref{eq03}). We numerically integrate Hamiltonian equations for systems with a single wave and with two waves. Figures \ref{fig2}(a,b) show electron motion in the energy/pitch-angle space due to the resonance with a single wave. Solid curves are resonant curves of $-\omega_0I_0-\omega_1I_1+m_ec^2\gamma={\rm const}$ for the wave frequency $\omega_0$ and for the wave frequency $\omega_1$. Electrons move along this curve due to phase bunching (small negative jumps of energy and pitch-angle; see bottom panels) and phase trapping (rare large positive jumps of energy and pitch-angle; see bottom panels). Conservation of $-\omega_0I_0-\omega_1I_1+m_ec^2\gamma$ and one of the momenta ($I_0$ or $I_1$) makes electron dynamics 1D in the energy/pitch-angle space. However, this dynamics becomes 2D in the system with two waves, when both $I_0$ and $I_1$ change, see Fig. \ref{fig2}(c). The electron moves along resonance curves and jumps between these curves due to $I_i$ jumps. There are still the same energy and pitch-angle jumps due to phase bunching and phase trapping (see bottom panels), but electron phase trajectory covers the entire energy/pitch-angle space. We describe this 2D dynamics with the mapping technique in this study. 

\subsection{Oblique whistler-mode wave}
\begin{figure}
 \centerline{\includegraphics[width=1\textwidth]{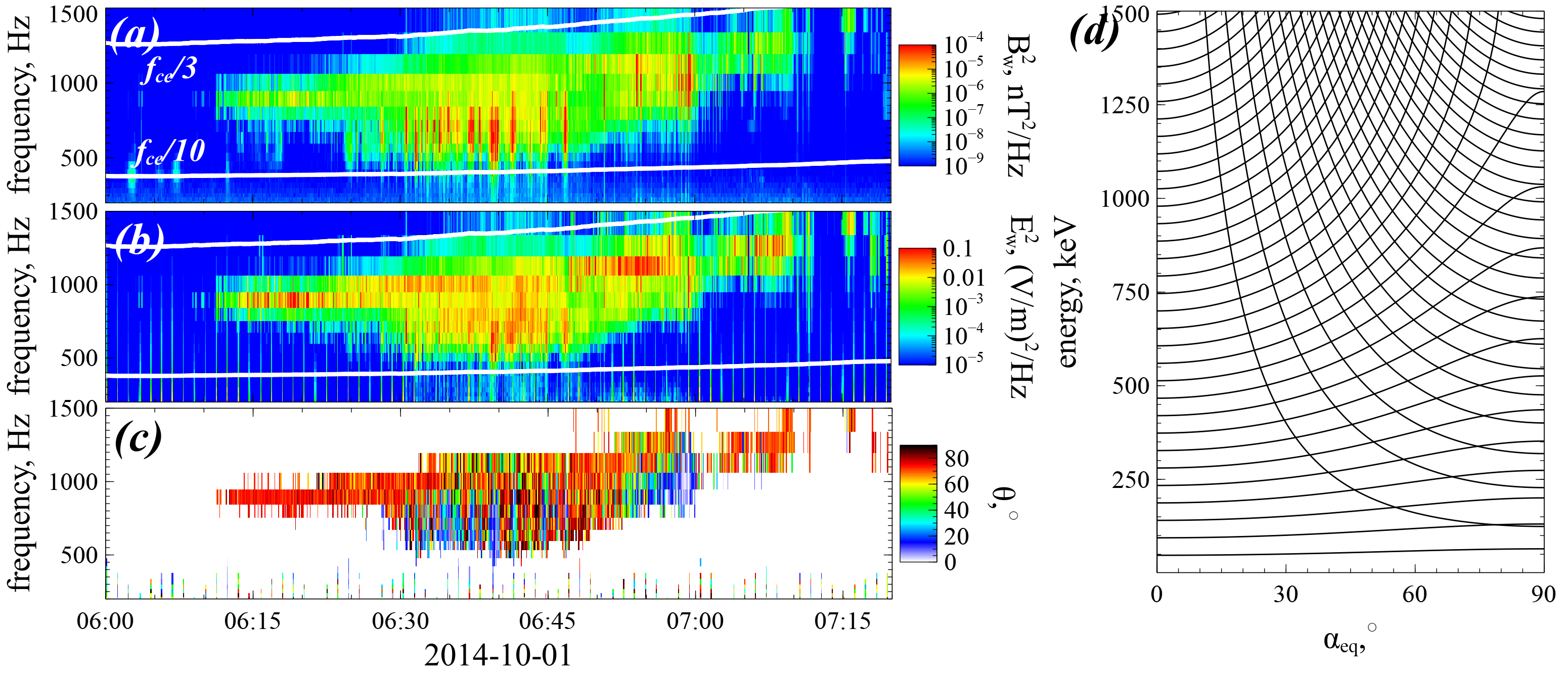}}
  \caption{Observational example of oblique whistler-mode wave by Van Allen Probe A \cite{Mauk13}: magnetic field spectrum (\textit{a}) and electric field spectrum (\textit{b}) are obtained from EMFISIS measurements \cite{Kletzing13}, wave-normal angle (\textit{c}) is estimated using using the singular value decomposition method \cite{Santolik03}. Resonance curves for the system with two observed whistler-mode waves (\textit{d}).}
\label{fig3}
\end{figure}

The second example corresponds to electron resonant interaction with a single oblique ($\theta\ne 0$) wave, for which Hamiltonian (\ref{eq01}) takes the form
\begin{eqnarray}
 H &=& m_e c^2 \gamma  + \frac{{eB_w }}{{k\gamma }}h_0 \sin \left( \phi  \right) + \sqrt {\frac{{2I_x \Omega _{ce} }}{{m_e c^2 }}} h_1 \frac{{eB_w }}{{k\gamma }}\sin \left( {\phi  + \psi } \right) \nonumber \\ 
 h_0  &=&  - \sqrt {\frac{{2I_x \Omega _{ce} }}{{m_e c^2 }}} C_1 J_1  + \left( {\frac{{p_\parallel }}{{m_e c}} + C_2 } \right)J_0 \sin \theta  \label{eq06}\\ 
 h_1  &=& \frac{1}{2}\left( J_2{\left( {\cos \theta  + C_1 } \right)  + J_0 \left( {\cos \theta  - C_1 } \right)} \right) + \left( {\frac{{p_\parallel }}{{m_e c}} + C_2 } \right)\frac{{kc}}{{2\Omega _{ce} }}\left( {J_2  + J_0 } \right) \sin \theta \nonumber  
 \end{eqnarray}
where we restrict our consideration to the  first two resonances: Landau resonance $n=0$ and the first cyclotron resonance $n=1$. The Bessel function argument is $\sqrt{2I_xk^2/m_e\Omega_{ce}}\sin\theta$. Such oblique whistler-mode waves are widely observed in the radiation belts \citep{Agapitov13:jgr, Agapitov15:jgr, Li16:statistics}, and their amplitudes are often sufficiently high for nonlinear resonances \citep{Agapitov15:grl:acceleration, Artemyev16:ssr, Mourenas16}. Figure \ref{fig3} shows an example of oblique whistler-mode wave measured by THEMIS spacecraft in the outer radiation belt. Electric and magnetic field spectra show the one wave power maximum around $f/f_{ce}\sim 1/5$ (see panels (a)\&(b)), i.e. this is a single wave. Wave normal angle $\theta\approx 70^\circ$ (see panel (c)), i.e., this wave propagates obliquely to the background magnetic field.

Using the same approach as the one we applied for Hamiltonian (\ref{eq03}), we introduce wave phases as new variables, $\varphi_0=\phi_0$ and $\varphi_1=\phi_1+\psi$, using the generating function \citep{Neishtadt06, Neishtadt14:rms}:
\begin{equation}
W = sP + \left( {\int {k(\tilde s)d\tilde s}  - \omega t} \right)I_0  + \left( {\int {k(\tilde s)d\tilde s}  - \omega t + \psi } \right)I_1 
\label{eq07}
\end{equation}
This function gives the new variables: $P=p-k_0I_0-k_1I_1$, $S=s$ (we keep $s$ notation), $I_x=I_1$, and new Hamiltonian $H_I=H+\partial W/\partial t=H-\omega I_0-\omega I_1$
\begin{eqnarray}
 H_I  &=&  - \omega I_0  - \omega I_1  + m_e c^2 \gamma  + \frac{{eB_w h_0 }}{{k\gamma }}\sin \varphi _0 
  + \sqrt {\frac{{2I_1 \Omega _{ce} }}{{m_e c^2 }}} \frac{{eB_w h_1 }}{{k\gamma }}\sin \varphi _1  \label{eq08}\\ 
 \gamma  &=& \sqrt {1 + \frac{{\left( {P + k_0 I_0  + k_1 I_1 } \right)^2 }}{{m_e^2 c^2 }} + \frac{{2I_1 \Omega _{ce} }}{{m_e c^2 }}}  \nonumber
 \end{eqnarray}
The resonance curves in the energy/pitch-angle space are given by two equations: $m_ec^2\gamma-\omega I_1={\rm  const} $ with $I_1=I_x=(\gamma^2-1)\sin^2\alpha_{eq}/2$ for the cyclotron resonance, and $I_x={\rm const}$ for the Landau resonance. Figure \ref{fig3}(d) shows that at $\alpha_{eq}<\pi/4$ Landau resonance curves cross cyclotron resonance curves almost transversely, i.e., in the Landau resonance electrons quickly change energy with weaker pitch-angle change, whereas in the cyclotron resonance the energy change is more effective than the pitch-angle change. 

\begin{figure}
 \centerline{\includegraphics[width=1\textwidth]{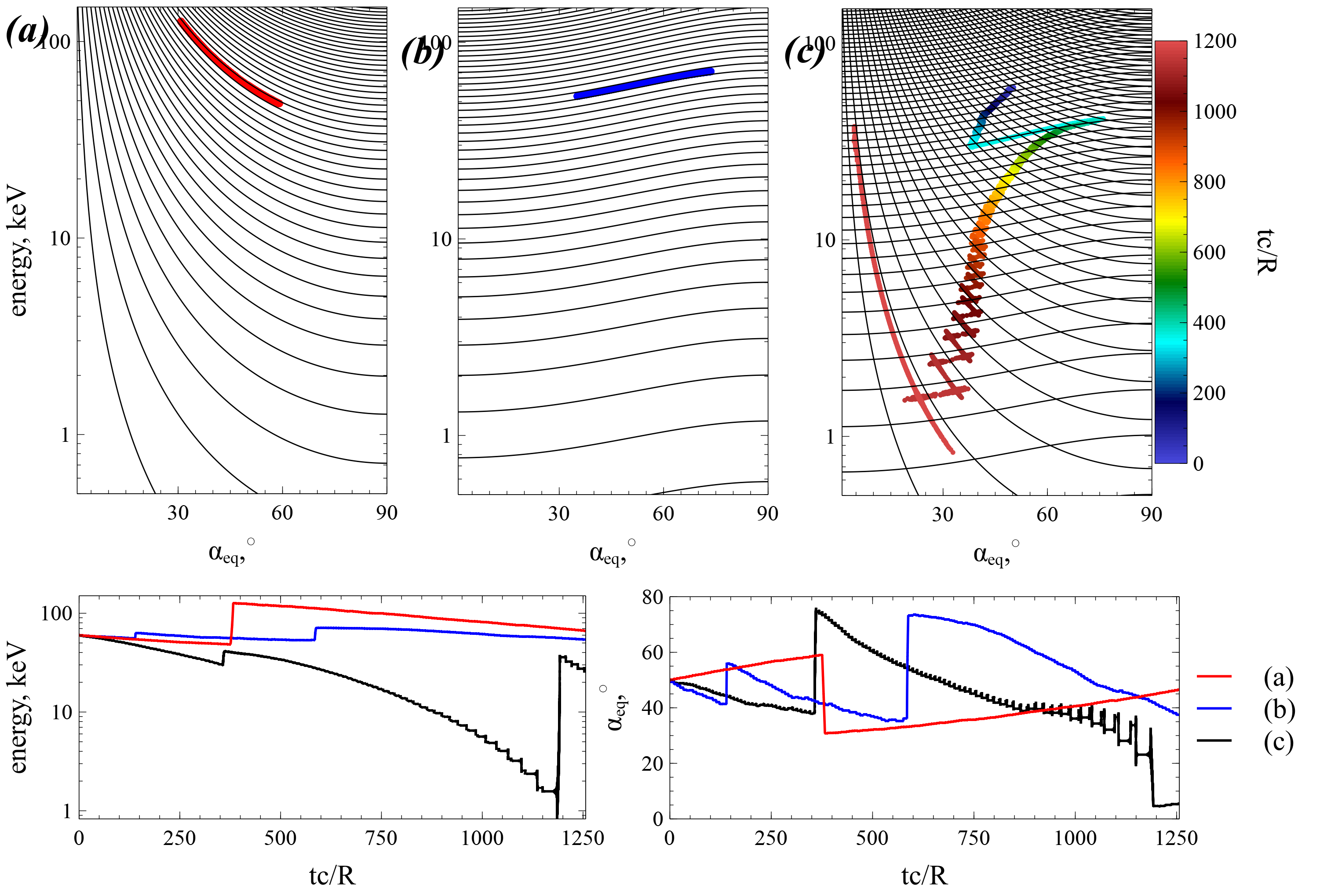}}
  \caption{Top panels show resonance curves (black) and electron trajectories in the energy/pitch-angle space for Hamiltonian (\ref{eq06}): only Landau resonance (\textit{a}), only cyclotron resonance   (\textit{b}), both resonances  (\textit{c}). In panel  (\textit{c}) color shows time. Bottom panels show energy and pitch-angle time series for three trajectories from top panels. System parameters are the same as in Fig. \ref{fig2}, wave frequency is $\omega=0.35\Omega_{ce}$, wave amplitude is $500$ pT, and wave normal angle $\theta$ is $5^\circ$ away of the resonance cone angle ${\rm acos}(\omega/\Omega_{ce})$ \citep[see discussion of such wave observations in][]{Cattell11:Wilson, Agapitov14:jgr:acceleration, Artemyev16:ssr, Mourenas16}. The wave amplitude distribution along magnetic field-lines is the same as one used in Fig. \ref{fig1}.}
\label{fig4}
\end{figure}

To demonstrate the effects of the two resonances on electron transport in energy/pitch-angle space, we numerically integrate Hamiltonian equations (\ref{eq08}) for three systems. Figure \ref{fig4}(a) shows results of the Landau resonance of the electron and oblique whistler-mode wave. The electron moves along a single resonant curve $I_x={rm const}$ with phase bunching responsible for pitch-angle increase and energy decrease, and the phase trapping responsible for pitch-angle decrease and energy increase (bottom panels). Figure \ref{fig4}(b) shows results of the cyclotron resonance: electron motion in the energy/pitch-angle space are quite similar to motions shown in Figs. \ref{fig2}(a\&b): phase bunching is responsible for pitch-angle and energy decrease, whereas the phase trapping is responsible for pitch-angle and energy increase (bottom panels). The combination of the two resonances results in rapid electron motion within the whole energy/pitch-angle domain, see Fig. \ref{fig4}(c). The phase bunching decreases electron energy in both resonances, but moves electron in opposite directions in pitch-angle. As a result, a resonant electron loses energy until it reaches the region with high probability of  trapping into the Landau resonance \citep{Artemyev13:pop}. After being trapped in Landau resonance, the electron gains energy and reaches the energy/pitch-angle domain where it can now be trapped into the cyclotron resonance with further energy increase. Such cycles of bunching, Landau trapping, and cyclotron trapping, quickly cover a large energy/pitch-angle domain for a single electron trajectory. 

\subsection{Field-aligned whistler-mode and EMIC waves}
\begin{figure}
 \centerline{\includegraphics[width=1\textwidth]{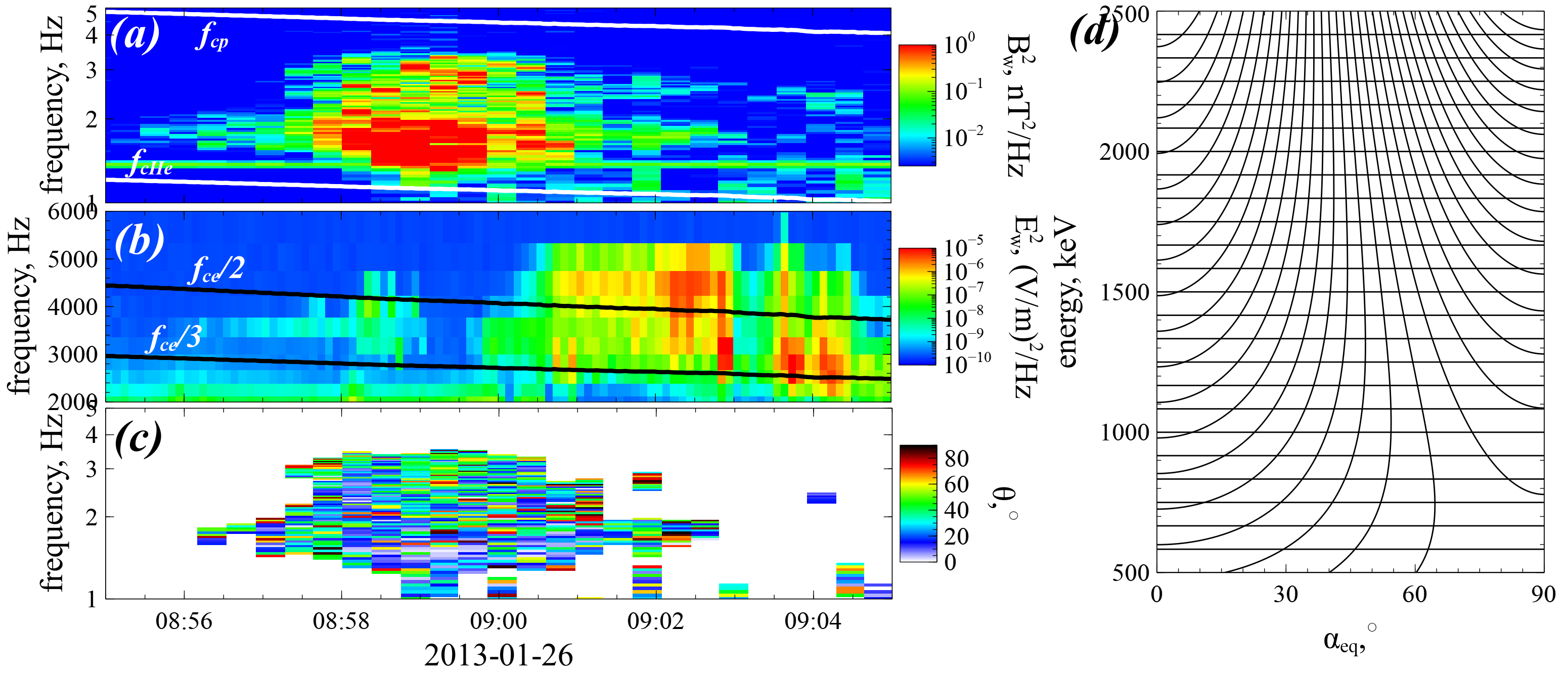}}
  \caption{Observational example of a system with whistler-mode and EMIC waves by Van Allen Probe A \cite{Mauk13}: magnetic field spectrum of EMIC wave (\textit{a}) and whistler-mode wave (\textit{b}) are obtained from EMFISIS measurements \cite{Kletzing13}, the wave-normal angle of the EMIC wave (\textit{c}) is estimated using the singular value decomposition method \cite{Santolik03}. Resonance curves for the system with the two observed whistler-mode waves (\textit{d}). }
\label{fig5}
\end{figure}

A third example is a system with field-aligned whistler-mode wave and field-aligned EMIC wave with polarization opposite  to the whistler-mode wave. The corresponding Hamiltonian of a relativistic electron (reduction of Hamiltonian (\ref{eq02})) takes the form
\begin{equation}
H = m_e c^2 \gamma  + \sqrt {\frac{{2I_x \Omega _{ce} }}{{m_e c^2 }}} \left( {\frac{{eB_{w,0} }}{{k_0 \gamma }}\sin \left( {\phi _0  + \psi } \right) + \frac{{eB_{w,1} }}{{k_1 \gamma }}\sin \left( {\phi _1  - \psi } \right)} \right)
\label{eq09}
\end{equation}
where $k_0=k_0(\omega_0,s)$ follows the whistler-mode wave dispersion, whereas $k_1=k_1(\omega_1,s)$ follows the EMIC wave dispersion. Figure \ref{fig5}(a\&b) shows a typical example of observation of such two waves: the high-frequency magnetic field spectrum shows the whistler-mode wave with $f/f_{ce}\sim f_{ce}/2$, whereas the low-frequency magnetic field spectrum shows the EMIC wave with $f/f_{cp}\sim f_{cp}/2$ ($f_{cp}$ is the proton gyrofrequency). The EMIC wave is field-aligned (see panel (c)). Due to the low EMIC wave frequency, the resonance condition $\dot\phi_1-\dot\psi=k_1p/\gamma-\omega_1-\Omega_{ce}/\gamma=0$ can be reduced to $k_1p=\Omega_{ce}$, with typical $k_1$  about the inverse ion inertial length \citep{Silin11}. Thus, only high-energy electrons (with large enough $p$) can resonate with EMIC waves (e.g., in the Earth radiation belts the resonant energy is typically larger than $\sim 1$ MeV, see \citet{Thorne&Kennel71, Summers&Thorne03, Shprits16,Chen19}). Let us compare whistler-mode and EMIC wave resonance curves for such high energies.

\begin{figure}
 \centerline{\includegraphics[width=1\textwidth]{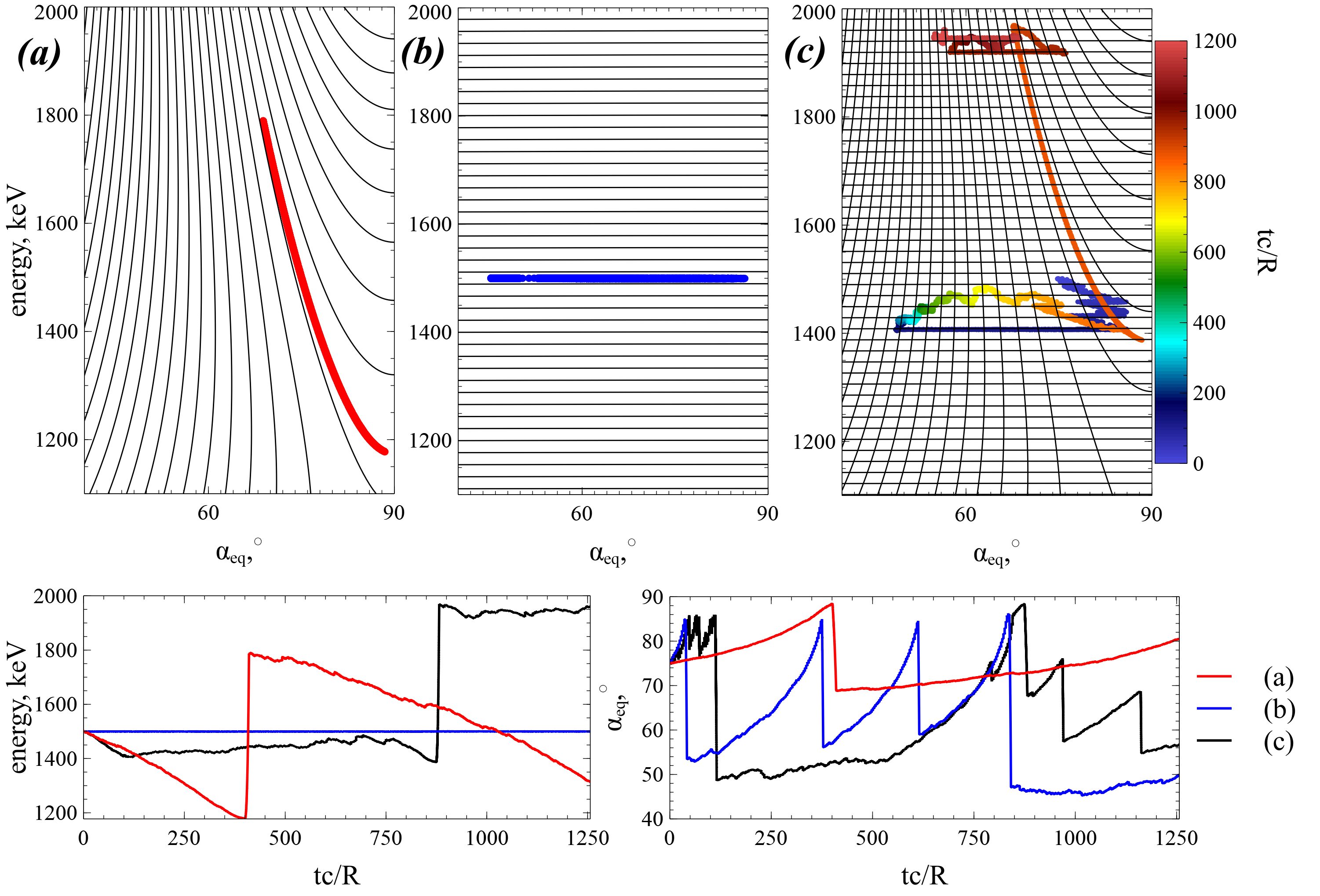}}
  \caption{Top panels show resonance curves (black) and electron trajectories in the energy/pitch-angle space for Hamiltonian (\ref{eq09}): only whistler-mode wave  (\textit{a}), only EMIC wave   (\textit{b}), both whistler-mode and EMIC waves  (\textit{c}). In panel  (\textit{c}) color shows time. Bottom panels show energy and pitch-angle time series for three trajectories in the top panels. System parameters are the same as in Fig. \ref{fig2}. EMIC wave is  $H^+$ band with the frequency $\omega=0.7\Omega_{cp}$ and amplitude $1$ nT \citep[see discussion of such wave observations in][]{Zhang16:grl, Kersten14}. Whistler-mode wave frequency is $\omega=0.35\Omega_{ce}$, and wave amplitude is $300$ pT.  The wave amplitude distribution along magnetic field-lines is the same as one used in Fig. \ref{fig1}.}
\label{fig6}
\end{figure}

First, we introduce wave phases as new variables, $\varphi_0=\phi_0+\psi$ and $\varphi_1=\phi_1-\psi$, with the generating function \citep{Neishtadt06, Neishtadt14:rms}:
\begin{equation}
W = sP + \left( {\int {k_0 (\tilde s)d\tilde s}  - \omega _0 t + \psi } \right)I_0  + \left( {\int {k_1 (\tilde s)d\tilde s}  - \omega _1 t - \psi } \right)I_1 
\label{eq10}
\end{equation}
This function gives new variables: $P=p-k_0I_0-k_1I_1$, $S=s$ (we keep $s$ notation), $I_x=I_0-I_1$, and new Hamiltonian $H_I=H+\partial W/\partial t=H-\omega I_0-\omega I_1$
\begin{eqnarray}
 H_I  &=&  - \omega _1 I_0  - \omega _1 I_1  + m_e c^2 \gamma  + \sqrt {\frac{{2\left( {I_0  - I_1 } \right)\Omega _{ce} }}{{m_e c^2 }}} \sum\limits_{i = 0,1} {\frac{{eB_{w,i} }}{{k_i \gamma }}\sin \varphi _i }  \nonumber \\ 
 \gamma  &=& \sqrt {1 + \frac{{\left( {P + k_0 I_0  + k_1 I_1 } \right)^2 }}{{m_e^2 c^2 }} + \frac{{2\left( {I_0  - I_1 } \right)\Omega _{ce} }}{{m_e c^2 }}}  \label{eq11}
\end{eqnarray}
The EMIC resonance curves are given by equation $m_ec^2\gamma-\omega_1I_1={\rm const}$, and taking into account the smallness of $\omega_1$ we obtain $\gamma\approx 0$, i.e. resonance curves are almost straight lines parallel to the energy axis (see Fig. \ref{fig5}(d)). The whistler-mode resonance curves ($m_ec^2\gamma-\omega_0I_0={\rm const}$ with $I_0=I_x+{\rm const}$) cross these lines: the EMIC wave is responsible for electron transport along pitch-angle space, and the whistler-mode wave leads to both pitch-angle and energy changes. Figure \ref{fig6}(a\&b) confirms this scenario: the EMIC wave resonates with small pitch-angle (large $p$) electron and phase bunch it to larger pitch-angles (phase trapping by EMIC waves is responsible for pitch-angle decrease; see bottom panel) with an approximate conservation of energy, whereas the whistler-mode wave can resonate with large pitch-angle electrons and transport them to smaller pitch-angles via phase bunching with energy decrease (moving them away from the EMIC wave resonance). 

The combination of EMIC and whistler-mode wave resonances (see Fig. \ref{fig6}(c)) can result in a very effective transport of large pitch-angle electrons to small pitch-angles (rapid electron losses): bunching of  $\sim 2$ MeV electrons with initially large pitch-angles results in electron transfer to small pitch-angles, where even faster EMIC phase trapping may move this electron to the loss-cone (see discussions of similar effects of combined EMIC and whistler-mode waves in the diffusive approximation in \citep{Mourenas16:grl,Zhang17}). From small pitch-angles (note that the loss-cone is not included in our simulations) the EMIC wave can transport an electron via phase bunching to higher pitch-angles, where whistler-mode resonance can accelerate it via trapping. As a result of so different resonant interactions with EMIC and whistler-mode waves, the electron trajectory can quickly fill up a large domain in the energy/pitch-angle space.  

\section{Mapping technique for multi-resonances}
\begin{figure}
\centering
 \centerline{\includegraphics[width=0.85\textwidth]{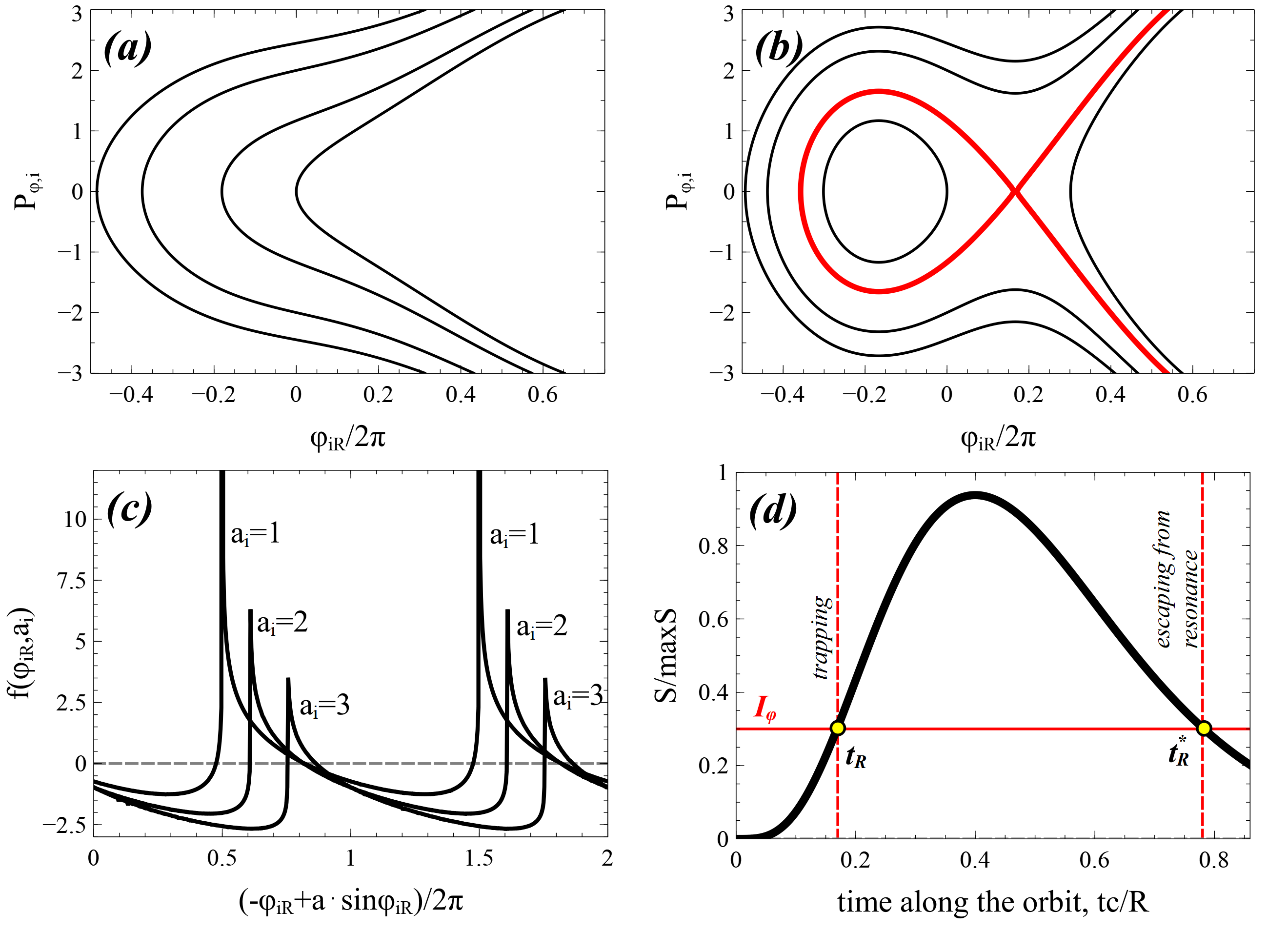}}
  \caption{Phase portraits of $F_{\varphi i}$ for systems with $|u_{iR}|<|\{\Lambda_{i}, I_{iR}\}|$ (\textit{a}) and with $|u_{iR}|>|\{\Lambda_{i}, I_{iR}\}|$ (\textit{b}). Black curves show contours of $F_{\varphi,i}={\rm const}$, red curve is the separatrix demarcating region of trapped (closed) electrons  orbits and region of transient (open) orbits.  
  Function $f(h_\phi,a)$ for several $a$ and $(-\varphi_{iR}+a\sin\varphi_{iR})/2\pi$ (\textit{c}) and scheme of trapping/detrapping (\textit{d}).}
\label{fig7}
\end{figure}

To describe the long-term evolution of electron dynamics in the energy/pitch-angle space, we propose to develop a map providing relations for each resonant interaction $\Delta\gamma=\Delta\gamma(\gamma,\alpha_{eq})$, $\Delta\alpha_{eq}=\Delta\alpha(\gamma,\alpha_{eq})$. Changes $\Delta\gamma, \Delta\alpha_{eq}$ are due to phase bunching (nonlinear scattering) and phase trapping. Thus, the first step in the construction of such a map is to derive equations for $\Delta\gamma, \Delta\alpha_{eq}$ driven by both these processes. We start with Hamiltonian (\ref{eq05}) and follow the standard procedure of Hamiltonian expansion around the resonant $I_0, I_1$ values \citep{Neishtadt14:rms, Artemyev18:cnsns}, which are defined by equations $\partial H_I/\partial I_i=0$:
\begin{equation}
\frac{{k_i I_{iR} }}{{m_e c}} =  - \frac{{P + k_{i'} I_{i'} }}{{m_e c}} - \frac{{\Omega _{ce} }}{{k_i }} + \frac{1}{{\sqrt {\left( {k_i c/\omega _i } \right)^2  - 1} }}\sqrt {1 - \left( {\frac{{\Omega _{ce} }}{{k_i c}}} \right)^2  - 2\frac{{\Omega _{ce} }}{{k_i c}}\frac{{P + \left( {k_{i'}  - k_i } \right)I_{i'} }}{{m_e c}}} 
\label{eq12}
\end{equation}
where $i’=0$ for $i=1$ and $i’=1$ for $i=0$. Expansion of Hamiltonian (\ref{eq05}) around $I_i=I_{iR}$ gives
\begin{eqnarray}
 H_{Ii}  &\approx& \Lambda _i  + m_e c^2 \frac{1}{2}g_i \left( {I_i  - I_{iR} } \right)^2  + u_{iR} \sin \varphi _i  \nonumber\\ 
 \Lambda _i  &=& m_e c^2 \gamma _{iR}  - \left( {\omega _0 I_0  + \omega _1 I_1 } \right)_{I_i  = I_{iR} }  \label{eq13}\\ 
 \gamma _{iR} & =& \frac{{\left( {k_i c/\omega _i } \right)}}{{\sqrt {\left( {k_i c/\omega _i } \right)^2  - 1} }}\sqrt {1 - \left( {\frac{{\Omega _{ce} }}{{k_i c}}} \right)^2  - 2\frac{{\Omega _{ce} }}{{k_i c}}\frac{{P + \left( {k_{i'}  - k_i } \right)I_{i'} }}{{m_e c}}}  \nonumber\\ 
 u_{iR}  &=& \sqrt {\frac{{2\Omega _{ce} \left( {I_0  + I_1 } \right)_{I_i  = I_{iR} } }}{{m_e c^2 }}} \frac{e}{{\gamma _{iR} }}\frac{{B_{w,i} }}{{k_i }} \nonumber\\ 
 g_i & =& \left. {\frac{{\partial ^2 \gamma }}{{\partial I_i^2 }}} \right|_{I_i  = I_{iR} }  = \frac{{k_i^2 }}{{m_e^2 c^2 }} \nonumber 
 \end{eqnarray}
where $\varphi_i$  are fast variables and $I_{i}-I_{iR})$ and $(s, P)$ are slow variables (note that $\Lambda_i$ does not depend on fast variables). Next, we introduce new variables $P_{\varphi i}=I_i-I_{iR}$ with the generating function $Q_i=(I_i-I_{iR})\varphi_i+s\tilde{P}_i$. New Hamiltonians are
\begin{eqnarray}
 F_i  &=& \Lambda _i \left( {\tilde s,\tilde P} \right) + m_e c^2 \frac{1}{2}g_i P_{\varphi i}^2  + u_{iR} \sin \varphi _i  \label{eq14}\\ 
  &\approx& \Lambda _i \left( {s,P} \right) + \left\{ {\Lambda _i ,I_{iR} } \right\}_{s,P} \varphi _i  + m_e c^2 \frac{1}{2}g_i P_{\varphi i}^2  + u_{iR} \sin \varphi _i  \nonumber
 \end{eqnarray}
where $\tilde{s}=s-(\partial I_{iR}/\partial P)\varphi_i$, $\tilde{P}=p+(\partial I_{iR}/\partial s)\varphi_i$, $\{…\}$ are Poisson brackets, and we expand $\Lambda(\tilde{s},\tilde{P})$ over small $\partial I_{iR}/\partial s$, $\partial I_{iR}/\partial P$ terms. Hamiltonian $F_i$ is the sum of $\Lambda_i(s,P)$ describing slow variable dynamics and pendulum Hamiltonian describing fast variable dynamics:
\begin{equation}
F_{\varphi i}  = m_e c^2 \frac{1}{2}g_i P_{\varphi i}^2  + \left\{ {\Lambda _i ,I_{iR} } \right\}_{s,P} \varphi _i  + u_{iR} \sin \varphi _i 
\label{eq15}
\end{equation}
where the coefficients depend on the slow variables. Figure \ref{fig7} shows phase portraits of $F_{\varphi i}$ for systems with $a_i=|u_{iR}/\{\Lambda_{i}, I_{iR}\}|<1$ (panel a) and with $a_i=|u_{iR}/\{\Lambda_{i}, I_{iR}\}|>1$ (panel b). For low wave amplitude $a_i<1$ the phase portrait does not contain closed orbits, i.e., all particles cross the resonance $\dot\varphi_i=m_ec^2g_iP_{\varphi,i}=0$ within an interval of about one period of $\varphi_i$. There are only weak scatterings in this regime with zero mean changes of $I_i$, and such scatterings can be described by the quasi-linear diffusion model for inhomogeneous plasma \citep[e.g.,][]{Karpman74:ssr, Albert10, Grach&Demekhov20}. For sufficiently high wave amplitude $a_i>1$, however, the phase portrait contains both closed and open orbits, i.e., there are now phase trapped particles oscillating around the resonance $\dot\varphi_i=m_ec^2g_iP_{\varphi,i}=0$ for a long time. Scattering (crossing of the resonance along the open orbits) would result in phase bunching with a small, yet nonzero mean change of $I_i$ \citep[see reviews by][and references therein]{Shklyar09:review, Albert13:AGU}, whereas phase trapping would significantly change $I_i$. We would like to include this nonlinear regime of wave-particle interaction into the map in energy/pitch-angle space. For this reason, we derive expressions for changes of $I_i$ due to phase bunching, $\Delta_{scat} I_i$, and due to phase trapping $\Delta_{trap} I_i$. As $\Delta_{scat} I_i$ is local, i.e. depends on particle and system characteristics at the resonance, we can keep slow variables unchanged for $\Delta_{scat} I_i$ evaluations:
\begin{eqnarray}
 \Delta _{scat} I_i  &=& 2\int\limits_{ - \infty }^{t_{iR} } {\frac{{\partial H_{Ii} }}{{\partial I}}dt}  = \frac{{2u_{iR} }}{{m_e c^2 g_i }}\int\limits_{ - \infty }^{\varphi _{iR} } {\frac{{\cos \varphi _i d\varphi _i }}{{P_{\varphi i} }}} \nonumber \\ 
 &=& \sqrt {\frac{{2u_{iR} }}{{m_e c^2 g_i }}} \int\limits_{ - \infty }^{\varphi _{iR} } {\frac{{\sqrt {u_{iR} } \cos \varphi _i d\varphi _i }}{{\sqrt {F_{\varphi i}  - \left\{ {\Lambda _i ,I_{iR} } \right\}_{s,P} \varphi _i  - u_{iR} \sin \varphi _i } }}} \nonumber \\ 
  &=& \sqrt {\frac{{2u_{iR} }}{{m_e c^2 g_i }}} \int\limits_{ - \infty }^{\varphi _{iR} } {\frac{{\sqrt {u_{iR} } \cos \varphi _i d\varphi _i }}{{\sqrt {\left\{ {\Lambda _i ,I_{iR} } \right\}_{s,P} \left( {\varphi _{Ri}  - \varphi _i } \right) + u_{iR} \left( {\sin \varphi _{Ri}  - \sin \varphi _i } \right)} }}} \label{eq16}\\ 
  &=& \sqrt {\frac{{2u_{iR} }}{{m_e c^2 g_i }}} \int\limits_{ - \infty }^{\varphi _{iR} } {\frac{{\sqrt{a_i} \cos \varphi _i d\varphi _i }}{{\sqrt {\left( {\varphi _{Ri}  - \varphi _i } \right) + a_i \left( {\sin \varphi _{Ri}  - \sin \varphi _i } \right)} }}} = \sqrt {\frac{{2u_{iR} }}{{m_e c^2 g_i }}} f_i \left( {\varphi _{Ri} ,a_i } \right) \nonumber
 \end{eqnarray}
where $t_{iR}$ is the time of passage through the resonance, $\varphi_{iR}$ is the wave phase at this time, and we use $\dot\varphi_{i}=m_ec^2g_iP_{\varphi i}=2^{1/2}\sqrt{F_{\varphi i}-\{\Lambda_{i}, I_{iR}\}\varphi_i-u_{Ri}\sin\varphi_i}$, $F_{\varphi i}=\{\Lambda_{i}, I_{iR}\}\varphi_{Ri}+u_{Ri}\sin\varphi_{Ri}$ (resonant energy $F_{\varphi i}$ value evaluated at $P_{\varphi i}=0$). Note Eq. (\ref{eq16}) describes $\Delta_{scat}I_i$ change for the particle motion thought the resonance from $-\infty$ to resonant $\varphi_{iR}$, whereas the motion in opposite direction would result in change of sign of $\Delta_{scat}I_i$. Function $f_i(a_i,\varphi_{iR})$ is periodic for $\varphi_{iR}$, see Fig. \ref{fig7}(c). Although the sign of $f_i$ changes within one $\varphi_{iR}$ period, the mean value of this function for $a_i>1$ is not zero, providing the effect of phase bunching. To consider the precise $\Delta_{scat} I_i$ dependence on $\varphi_{iR}$ in the mapping, one would need to keep information about resonant phase $\varphi_i$ and calculate the phase gain between resonances. However, the phase is fast rotating, and even a small change of $\varphi_i$ at the resonance would result in a significant change of phase gain between resonances. Therefore, we can assume that $\varphi_{iR}$ is a random variable with a uniform distribution of the resonant energy $F_{\varphi i}(\varphi_{iR})$ at $P_{\varphi i}=0$ axis (see justification of this assumption in \citet{Itin00, Artemyev20:rcd, Artemyev20:pop}), and all resonant particles with the same slow variables (same energy and pitch-angle) at the resonance would experience the same $\Delta I_i$ change equal to $\langle \Delta I_i \rangle$ averaged over the resonant energy \citep{Artemyev20:pop}.   

An important property of $f$ function from Eq. (\ref{eq16}) is that being averaged over energies in resonance, $F_{\varphi i}=\{\Lambda_{i}, I_{iR}\}\varphi_{Ri}+u_{Ri}\sin\varphi_{Ri}$, this function gives
\begin{equation}
\left\langle {f_i } \right\rangle  =-
\sqrt {\frac{m_ec^2g_i}{2u_{iR}} }\frac{S}{{2\pi }}=
 -\sqrt{\frac{8|u_{iR}|}{a_im_ec^2g_i}}\int\limits_{ \varphi_{i-} }^{\varphi _{iR} } {\sqrt {\left( {\varphi _{Ri}  - \varphi _i } \right) + a_i \left( {\sin \varphi _{Ri}  - \sin \varphi _i } \right)} d\varphi _i } \label{eq17}
\end{equation}
where $S$ is the area surrounded by the separatrix in the phase portrait in Fig. \ref{fig7}(b) (see details of Eq. (\ref{eq17}) derivations in \citet{Neishtadt99} and \citet{Artemyev18:cnsns}). Therefore, the $\Delta_{scat}I_i$ change due to phase bunching is equal to $-S/2\pi$ and for $a_i\gg 1$ (i.e. for very weak magnetic field inhomogeneity; note $\{\Lambda_i, I_{iR}\}\sim \partial/\partial s$) we have $\Delta_{scat}I_i=-8\sqrt{2u_{iR}/m_ec^2g_i}$ where $S=16\sqrt{2u_{iR}/m_ec^2g_i}$ is the width of the resonance for large amplitude waves \citep{Palmadesso72,Karimabadi90:waves}.

The change of $I_i$ due to phase bunching (nonlinear scattering) is sufficiently small to consider this process locally in energy/pitch-angle space, i.e., $\Delta_{scat}I_i\ll I_i$ (see discussion of exceptions for $\Delta_{scat}I_i\sim I_i$ in Appendix A), whereas the change of $I_i$ due to phase trapping is essentially non-local. To evaluate $\Delta_{trap}I_i$, we take into account that $I_i=I_{iR}$ in the resonance (during the trapping), the trapping time is defined as $2\pi I_\phi=\int{P_{\varphi i}d\varphi_{iR}}=S$, and $I_{\phi}$ is conserved during the trapping (because trapped particles oscillate in the $(\varphi_{iR}, P_{\phi i})$ plane much faster than the system evolves (much faster than variations of slow variables $s,P$). Thus, the trapping time is defined as the time of arrival to the resonance  $t_R$ with $\dot S(t_R)>0$ (the growth of the area surrounded by the separatrix allows trapping of particles moving along open trajectories into closed trajectories), whereas the time $t_R^*$ of escape from the trapping is defined by $S(t_R^*)=S(t_R)$ and $\dot S(t_R^*)<0$ (see scheme in Fig. \ref{fig7}(d)):
\begin{equation}
\Delta _{trap} I_i  = I_{iR} \left( {t_{iR}^* } \right) - I_{iR} \left( {t_{iR} } \right),\quad S\left( {t_{iR}^* } \right) = S\left( {t_{iR} } \right),\quad \dot S\left( {t_{iR} } \right) > 0,\quad \dot S\left( {t_{iR}^* } \right) < 0
\label{eq18}
\end{equation}
At the resonance, an electron can be scattered (i.e., experience the phase bunching) or trapped, and this depends on the $\varphi_{iR}$ value \citep[e.g.,][]{Albert93, Itin00, Grach&Demekhov18:I}. However, as $\varphi_{iR}$ is a fast oscillating variable, we can consider the so-called probability of trapping instead of tracing the precise   $\varphi_{iR}$ value: the range of $\varphi_{iR}$ of trapped particles, i.e., the ratio of trapped particles to the total number of resonant particles for a single resonance, is the probability of trapping, $\Pi_i$ \citep[e.g.,][and references therein]{bookAKN06}. For small $\Pi_i$, this probability is defined as the ratio of the change of the area under the separatrix, $\dot S$, and the total resonant flux $\int_0^{2\pi}\dot P_{\varphi,i}d\phi=2\pi \{\Lambda_i, I_{iR}\}$: $\Pi_i=\dot S/2\pi \{\Lambda_i, I_{iR}\}=\{S, F_i\}/2\pi \{\Lambda_i, I_{iR}\}$. This definition of the trapping probability has been verified for various plasma systems \citep[e.g.,][]{Artemyev14:pop, Leoncini18, Vainchtein18:jgr}. Therefore, the resonant interaction can be characterized by $\Pi$, $\Delta_{trap} I_i$, and  $\Delta_{scat} I_i$.

Due to conservation of $H_i=m_ec^2\gamma-\omega_0I_0-\omega_1I_1$, changes of $I_i$ are directly related to $\gamma$ changes, whereas the $I_x=I_0+I_1$ relation gives the pitch-angle change:
\begin{eqnarray*}
 &&\frac{{\omega _i \sin ^2 \alpha _{eq} }}{{\gamma ^2  - 1}}\left( {\Delta _i \gamma } \right)^2  - 2\Delta _i \gamma \frac{{\Omega _{eq} - \gamma \omega _i \sin ^2 \alpha _{eq} }}{{\gamma ^2  - 1}} \\ 
  &+& \omega _i \left( {\sin ^2 \left( {\alpha _{eq}  + \Delta _i \alpha _{eq} } \right) - \sin ^2 \alpha _{eq} } \right) = 0 
\end{eqnarray*}
that for small changes (phase bunching) can be rewritten as
\begin{equation}
\Delta _i \alpha _{eq}  = \Delta _i \gamma \frac{{\Omega _{eq}  - \gamma \omega _i \sin ^2 \alpha _{eq} }}{{\omega _i \sin \alpha _{eq} \cos \alpha _{eq} \left( {\gamma ^2  - 1} \right)}}
\label{eq19}
\end{equation}

Therefore, the map for one resonance can be written as
\begin{equation}
\begin{array}{l}
 \left( {\begin{array}{*{20}c}
   {\bar \gamma }  \\
   {\bar \alpha _{eq} }  \\
\end{array}} \right) = \left( {\begin{array}{*{20}c}
   {G_{\gamma i} \left( {\gamma ,\alpha _{eq} } \right)}  \\
   {G_{\alpha i} \left( {\gamma ,\alpha _{eq} } \right)}  \\
\end{array}} \right) = \left( {\begin{array}{*{20}c}
   \gamma   \\
   {\alpha _{eq} }  \\
\end{array}} \right) + \left( {\begin{array}{*{20}c}
   {\Delta _i \gamma }  \\
   {\Delta _i \alpha _{eq} }  \\
\end{array}} \right) \\ 
 \Delta _i \gamma  = \omega _i \left\{ {\begin{array}{*{20}c}
   {\Delta _{scat} I_i \left( {\gamma ,\alpha _{eq} } \right),} & {\xi _i  \in \left[ {\Pi _i \left( {\gamma ,\alpha _{eq} } \right),1} \right]}  \\
   {\Delta _{trap} I_i \left( {\gamma ,\alpha _{eq} } \right),} & {\xi _i  \in \left[ {0,\Pi _i \left( {\gamma ,\alpha _{eq} } \right)} \right)}  \\
\end{array}} \right. \\ 
 \Delta _i \alpha _{eq}  = \left\{ {\begin{array}{*{20}c}
   {\frac{{\Omega _{eq}  - \gamma \omega _i \sin ^2 \alpha _{eq} }}{{\sin \alpha _{eq} \cos \alpha _{eq} \left( {\gamma ^2  - 1} \right)}}\Delta _{scat} I_i \left( {\gamma ,\alpha _{eq} } \right),} & {\xi _i  \in \left[ {\Pi _i \left( {\gamma ,\alpha _{eq} } \right),1} \right]}  \\
   {\Delta _i \alpha _{eq} \left( {\Delta _{trap} I_i ,\gamma ,\alpha _{eq} } \right),} & {\xi _i  \in \left[ {0,\Pi _i \left( {\gamma ,\alpha _{eq} } \right)} \right)}  \\
\end{array}} \right. \\ 
 \end{array}
\label{eq20}
\end{equation}
where $\xi_i$ is a random variable uniformly distributed in $[0,1]$. If there are two resonances (one with the first wave and another one with the second wave) during one electron bounce period $\tau_b$, then over this period the electron energy/pitch-angle change should be
\begin{equation}
\left( {\begin{array}{*{20}c}
   {\bar \gamma }  \\
   {\bar \alpha _{eq} }  \\
\end{array}} \right) = \left( {\begin{array}{*{20}c}
   {G_{\gamma 1} \left( {G_{\gamma 0} \left( {\gamma ,\alpha _{eq} } \right),G_{\alpha 0} \left( {\gamma ,\alpha _{eq} } \right)} \right)}  \\
   {G_{\alpha 1} \left( {G_{\gamma 0} \left( {\gamma ,\alpha _{eq} } \right),G_{\alpha 0} \left( {\gamma ,\alpha _{eq} } \right)} \right)}  \\
\end{array}} \right)\quad 
\label{eq21}
\end{equation}

\begin{figure}
 \centerline{\includegraphics[width=1.0\textwidth]{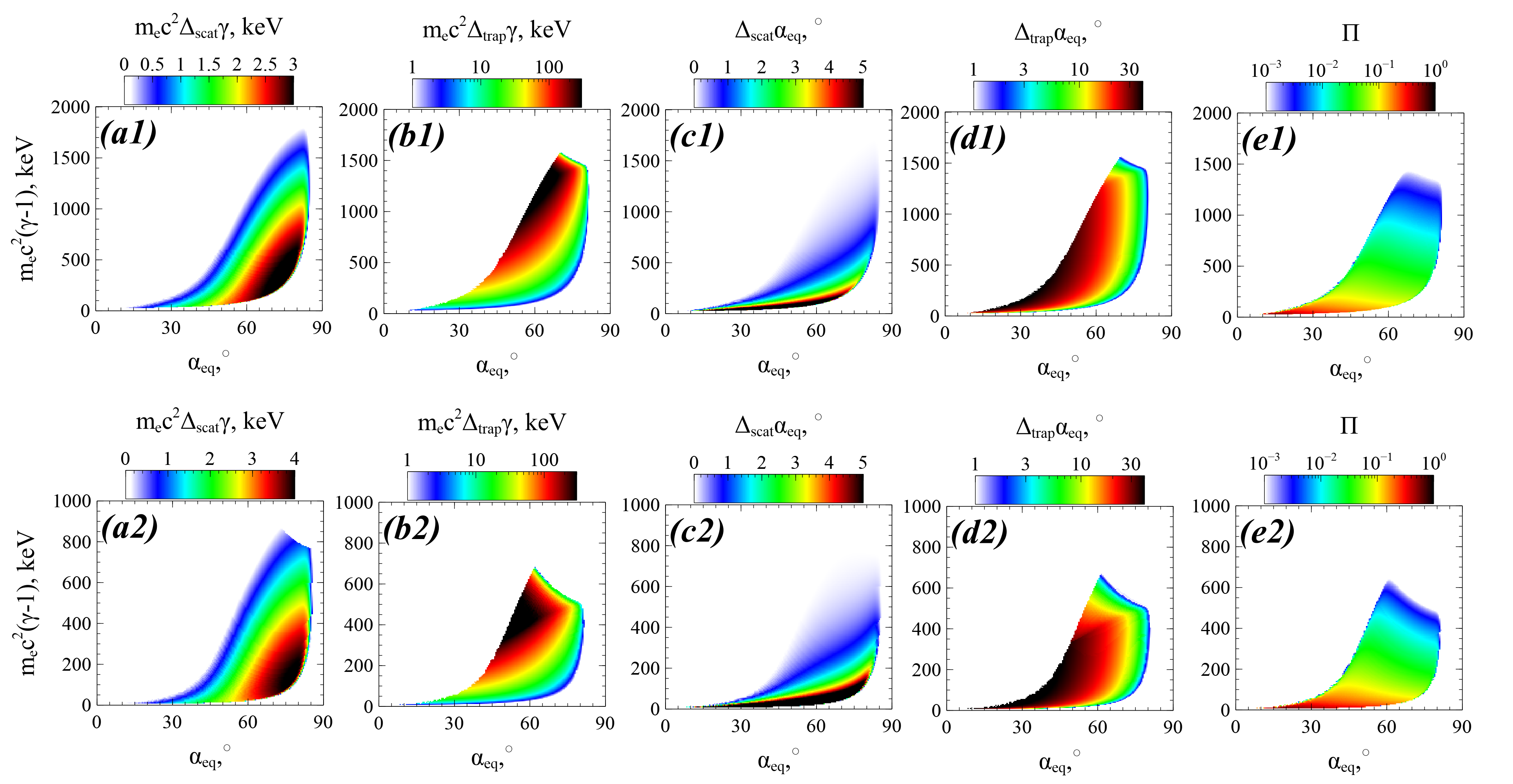}}
  \caption{System characteristics for two field-aligned whistler-mode waves with the parameters as in Fig. \ref{fig1}: energy change due to scattering (\textit{a}) and trapping (\textit{b}), pitch-angle change due to scattering (\textit{c}) and trapping (\textit{d}), trapping probability (\textit{e}).}
\label{fig8}
\end{figure}

Figure \ref{fig8} shows ten main characteristics of map (\ref{eq21}) in the energy/pitch-angle space: amplitudes of scattering $\Delta_{scat,i}\gamma$, $\Delta_{scat, i}\alpha_{eq}$, amplitudes of trapping $\Delta_{trap,i}\gamma$, $\Delta_{trap, i}\alpha_{eq}$, and trapping probabilities $\Pi_i$ for two field-aligned whistler-mode waves. To derive these characteristics for given energy and pitch-angle, we (1) calculate $\gamma$, $\alpha_{eq}$ and resonance location $s_R$ given by equation $I_i=I_{iR}$; (2) determine coefficients of Hamiltonian $F_i$, $S$, $\dot S$, and trapping probability $\Pi$ at $s_R$; (3) determine $\Delta_{scat}I_i=-S/2\pi$, position of escape from the resonance $s_R^*$ (if $\dot S(s_R)>0$), and $\Delta_{trap}I_i=I_{iR}(s_R^*)- I_{iR}(s_R)$; (4) recalculate $\Delta_{scat}I_i$, $\Delta_{trap}I_i$ into energy and pitch-angle changes. Numerical verification of this technique of $\Delta_{scat,i}\gamma$, $\Delta_{trap,i}\gamma$, $\Pi_i$ with test particle trajectories can be found in \citet{Vainchtein18:jgr, Artemyev20:pop}.

\begin{figure}
 \centerline{\includegraphics[width=1.0\textwidth]{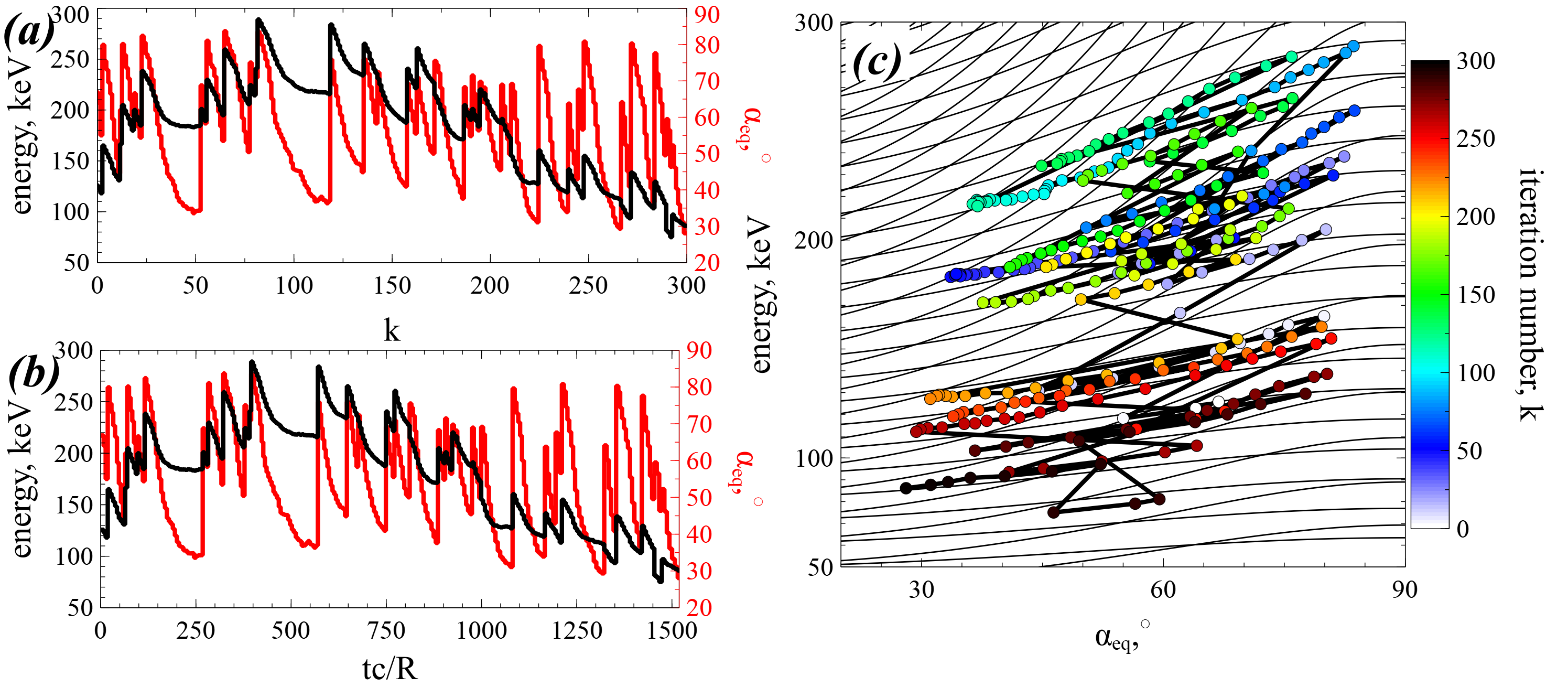}}
  \caption{A sample trajectory obtained using map (\ref{eq21}): energy and pitch-angle versus number of map iterations (\textit{a}), energy and pitch-angle versus time (\textit{b}), particle trajectory in the energy/pitch-angle space (\textit{c}).}
\label{fig9}
\end{figure}

Substituting characteristics from Figure \ref{fig8} into map (\ref{eq21}), we evaluate dynamics of resonant electrons. Figure \ref{fig9} shows a sample trajectory: energy and pitch-angle are plotted versus the number of iterations $k$ and versus time $t=\sum_k\tau_{b,k}(\gamma,\alpha_{eq})$. The trajectory obtained with the mapping technique contains all elements that can be found in the numerically integrated trajectory (compare with Fig. \ref{fig2}): energy decrease due to phase bunching and rare jumps due to phase trapping. Note that the bounce period is given by $\tau_b=4\int_0^{s_{\max}}ds/p$ with $p=m_ec^2\sqrt{1-\gamma-2I_x\Omega_{ce}(s)}$ and $2I_x\Omega_{ce}(s_{\max})=1-\gamma$. Any direct comparison of trajectories obtained via numerical integration and mapping technique is not possible due to significant randomization of resonant electron motion, i.e. trajectories in energy/pitch-angle plane for two test electrons can differ significantly even with small difference of initial electron phases \citep[e.g.,][]{Shklyar81,LeQueau&Roux87,Albert01}. Thus, the verification of map (\ref{eq21}) is mainly based on verification of Eqs. (\ref{eq16},\ref{eq18}) \citep[see][]{Artemyev15:pop:probability, Artemyev16:pop:letter,Vainchtein18:jgr} and on verification of 1D analogs of this map \citep[see][]{Artemyev20:pop}.

Using map (\ref{eq21}), we can simulate the evolution of the electron distribution function as an ensemble of test trajectories. We start with the test simulation of electron spread in the energy/pitch-angle space. Four populations of electrons with small ranges of initial energy and pitch-angles are traced for $500$ interactions and their positions in energy/pitch-angle space are shown at six different times, see Fig. \ref{fig10}. White color shows the area of resonant wave-particle interaction (see Appendix B for a definition of this area and for technical details of map (\ref{eq21}) application). Electrons of different initial populations quickly (already after $tc/R\sim 50$, i.e., $\sim 15$ resonant interactions) spread within a wide pitch-angle range, but are somehow separated in energy. After $tc/R\sim 300$ ($\sim 80$ resonant interactions) the populations fill large areas in energy/pitch-angle space and start overlapping. After $tc/R\sim 1000$ ($\sim 250$ resonant interactions) the entire energy/pitch-angle space is covered, and electrons from low energy populations (black and blue) reach high energies ($\sim 1$ MeV), whereas electrons from high-energy populations (red and magenta) decelerate with energy losses of several hundred keVs. Such fast phase mixing should result in spreading and smoothing of the electron phase space density.

\begin{figure}
\centerline{\includegraphics[width=1\textwidth]{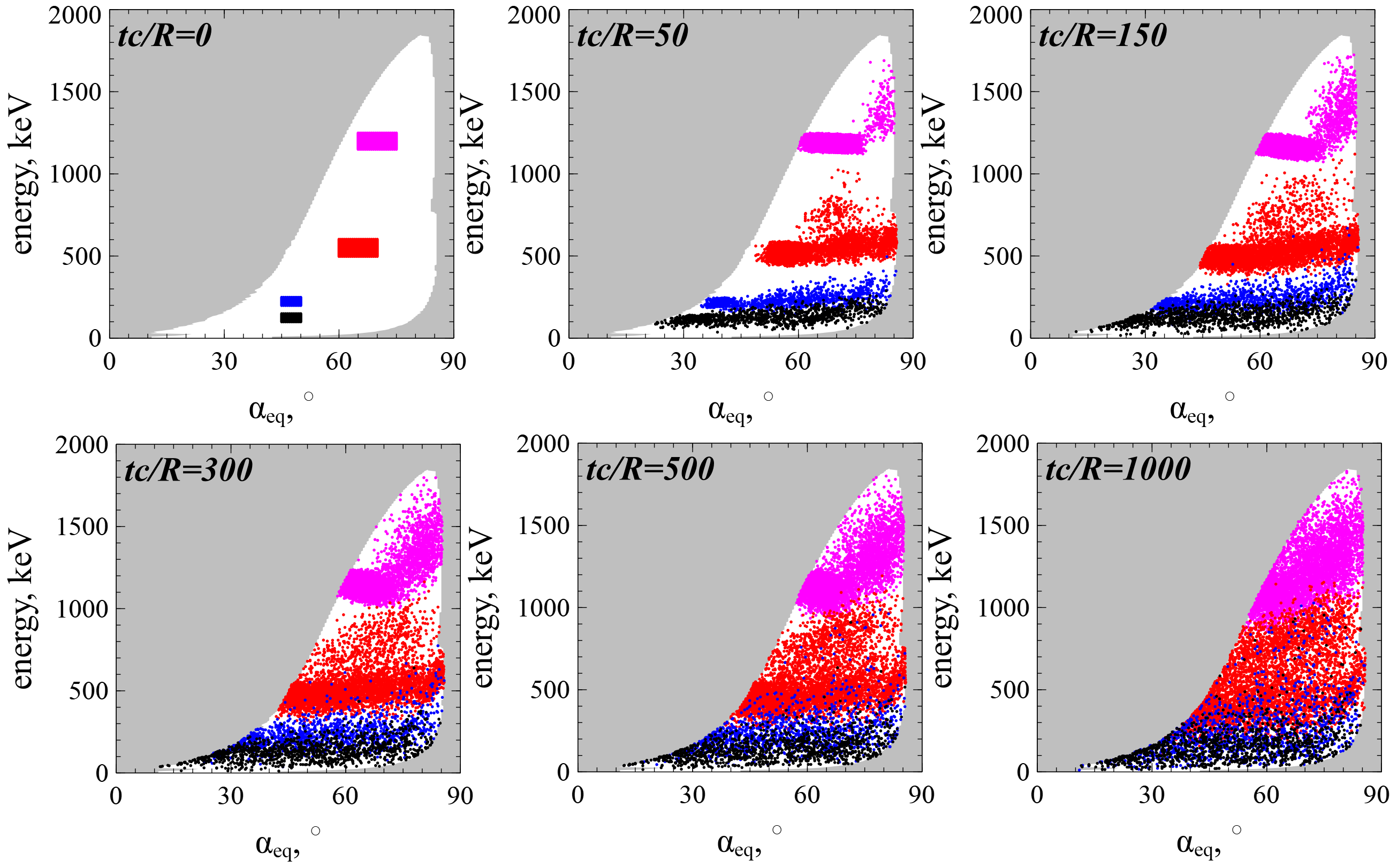}}
  \caption{Evolution of four electron populations modeled with map (\ref{eq21}). Six moments of time are shown. White color shows the area of wave-particle nonlinear resonant interaction, see Appendix B.}
\label{fig10}
\end{figure}

\begin{figure}
\centerline{\includegraphics[width=1\textwidth]{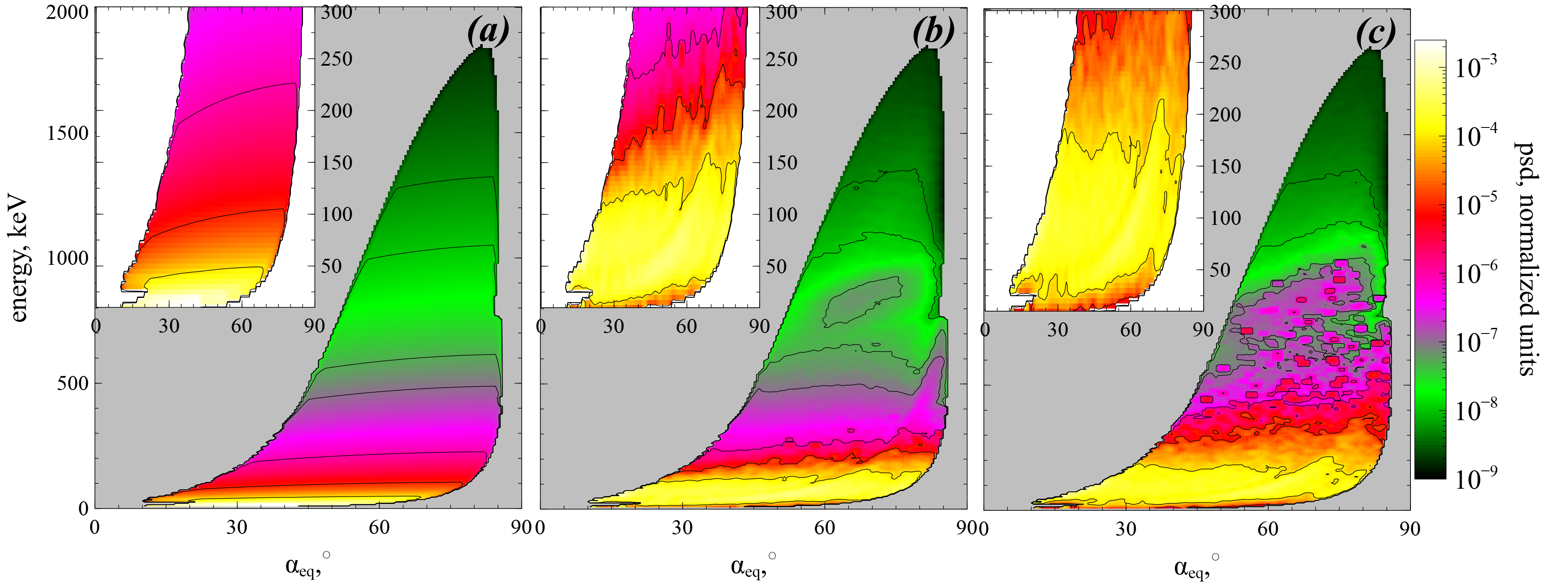}}
  \caption{Evolution of electron distribution modeled with map (\ref{eq21}). Three moments of time are shown: $tc/R=0$ (\textit{a}), $tc/R=300$ (\textit{b}), $tc/R=1000$ (\textit{c}). The initial distribution $f_0(\gamma,\alpha)=C\cdot\sin\alpha_{eq}\cdot(\gamma-1)^{-3}$ in panel (\textit{a}).}
\label{fig11}
\end{figure}

To examine the evolution of the electron phase space density, we start with a power law distribution $f_0(\gamma,\alpha)=C\cdot\sin\alpha_{eq}\cdot(\gamma-1)^{-3}$ typical in the radiation belts, and fit this distribution by $2\cdot 10^7$ trajectories. There are $180\times400$ pitch-angle/energy values, and $\sim 22600$ within the resonant area; for each value within the resonant area, we run $1000$ trajectories. Each trajectory is traced for 300 interactions with the map (\ref{eq21}), and corresponding $\alpha_{eq}(k)$, $\gamma(k)$ profiles transferred to time series. Then, we recalculate the distribution from $f_0$ using phase space density conservation along the trajectories. Figure \ref{fig11} shows three snapshots of the distribution $f(\alpha,\gamma)$ at different times (inserted panels show the low energy sub-interval). The rapid evolution of the distribution function results in phase space density flattening within the resonant region: there is an increase of high-energy/small pitch-angle phase space density and a decrease of low energy/large pitch-angle phase space density. During the simulation time, one electron can be trapped several times, i.e., most of particles circulate in the energy/pitch-angle space, because trappings bring them to the high energy region from which they then drift by bunching. Such a circulation also comprises successive trappings by two waves that bring electrons to the very high-energy region, whereas long periods of phase bunching without trappings can transport very energetic electrons to quite low energies. The last two phenomena are less frequent, and mixing of $\sim 1$ MeV electrons with $< 100$ keV electrons is slower than mixing within energy localized domains. 

\begin{figure}
\centerline{\includegraphics[width=0.5\textwidth]{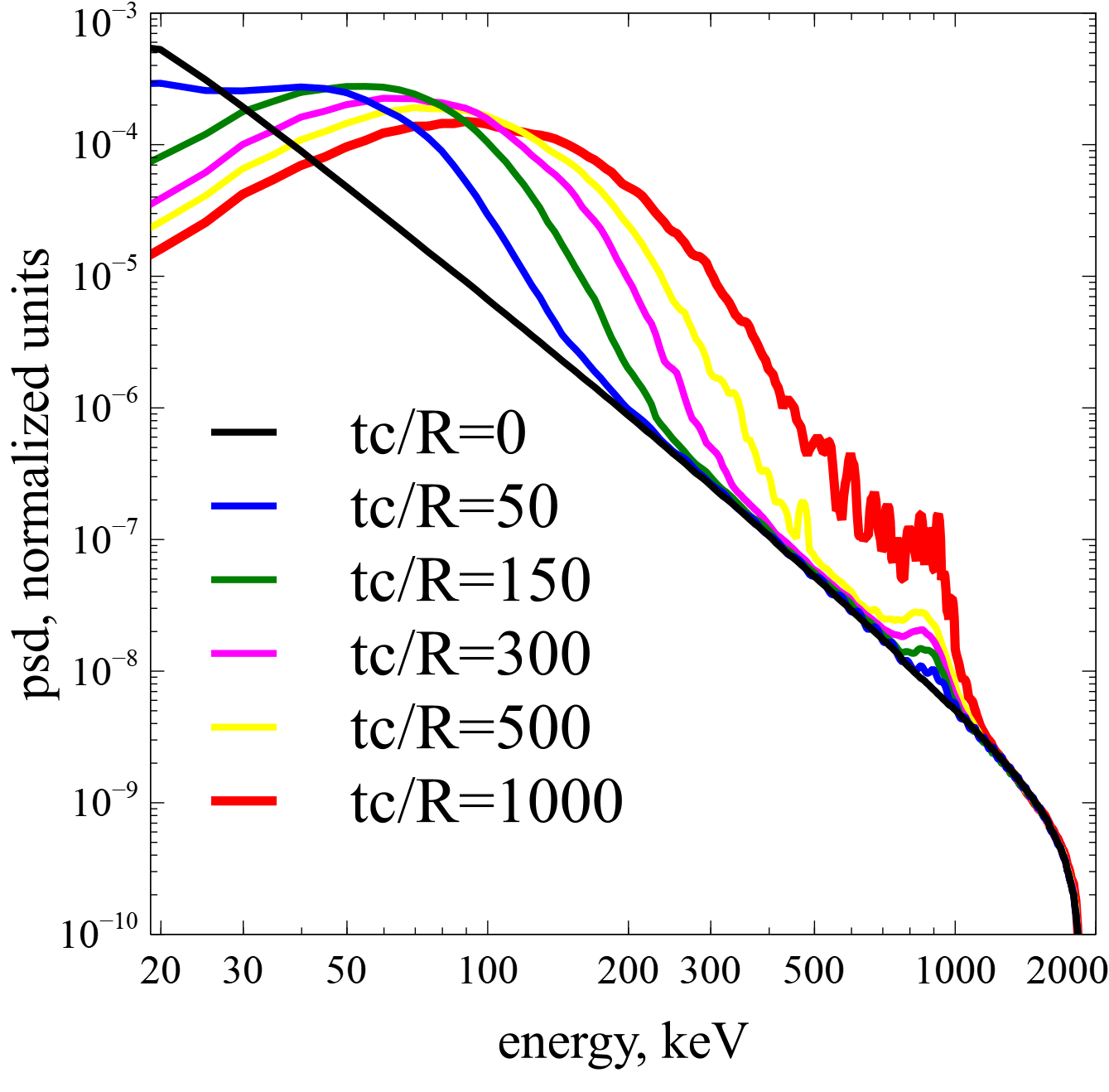}}
  \caption{Pitch-angle integrated electron distribution, $\int_0^{\pi}\sin\alpha_{eq}\cdot psd(E,\alpha_{eq})d\alpha_{eq}$, from Fig. \ref{fig11}.}
\label{fig12}
\end{figure}

The general trend of the resonant electron transport in the energy/pitch-angle space is the reduction of phase space density gradients. In the presence of a single wave, such a gradient smoothing occurs along the resonant curves, $\gamma-\omega_0I_0={\rm const}$ \citep{Artemyev20:pop}. In systems with two waves, the intersection of resonant curves $\gamma-\omega_0I_0={\rm const} $ and $\gamma-\omega_1I_1={\rm const}$ results in 2D gradient smoothing, i.e., we can expect a reduction of gradients in energy space after integration over pitch-angle. Figure \ref{fig12} shows such electron acceleration: increase of high-energy population and decrease of low energy population that result in gradient smoothing. This is the typical evolution of the electron distribution due to resonant interaction with whistler-mode waves (see similar results for nonlinear \citep{Vainchtein18:jgr} and quasi-linear \citep{Thorne13:nature, Li14:storm} simulations).

\section{Discussion and conclusions}
The proposed approach allows to investigate the long-term evolution of the electron distribution function in a system with nonlinear wave-particle interaction. This approach is based on the mapping technique that significantly simplifies electron trajectory integration by excluding from the consideration the main, adiabatic part of electron orbits and by focusing only on small intervals of resonant electron phase bunching and trapping. This approach is somewhat analogous to the Green function method proposed by \citep{Furuya08, Omura15} and to the nonlinear kinetic equation proposed by \citep{Artemyev16:pop:letter, Vainchtein18:jgr}. However, contrary to these other methods, the mapping does not require a very fine discretization of energy/pitch-angle space and it can easily be generalized to multi-wave systems. Resonances with different waves are very important for the destruction of the symmetry typical for the single wave system, where conservation of $(\gamma-\omega I)$ results in a reduced mixing in energy/pitch-angle space. Already, two waves with different characteristics are sufficient to produce a total mixing in energy/pitch-angle space (see Fig. \ref{fig10}) and a smoothing (reduction) of electron phase space density gradients (see Figure \ref{fig12}). The similar effect of fast mixing due to two independent resonances has been found in various dynamical systems with quite general properties \citep[e.g.][]{Gelfreich11,Itin&Neishtadt12}.

\begin{figure}
\centerline{\includegraphics[width=1.0\textwidth]{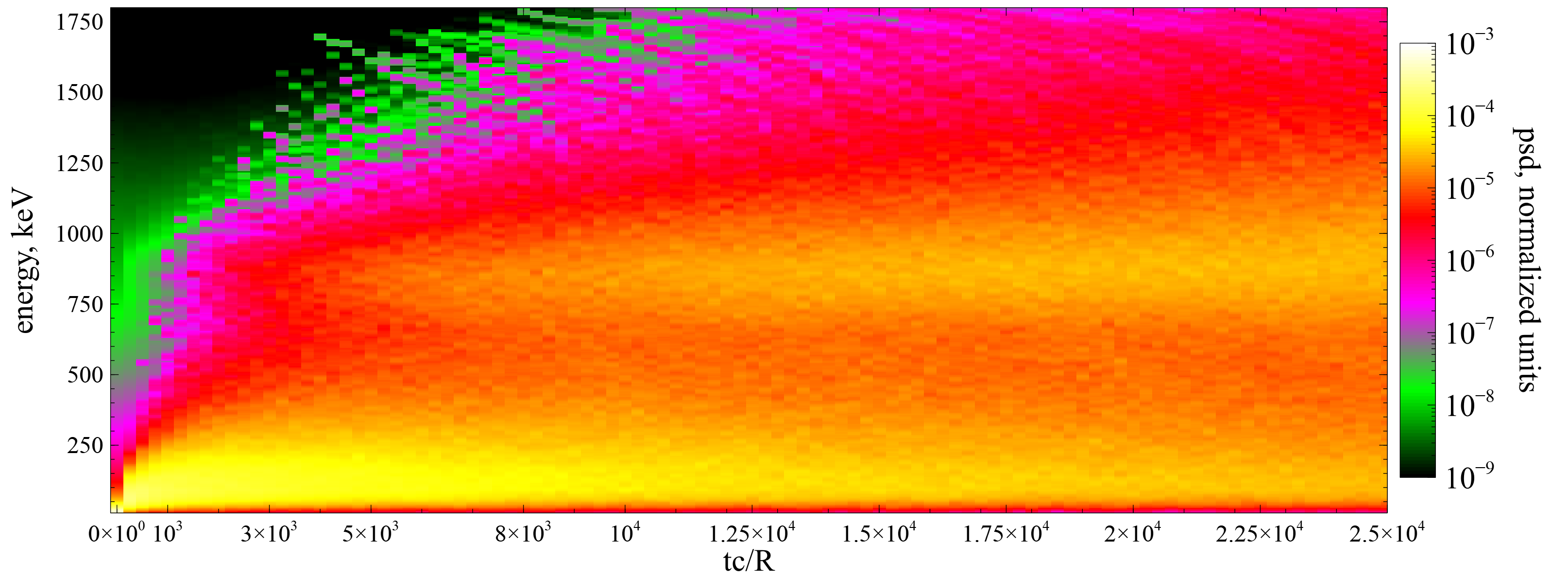}}
  \caption{Evolution of pitch-angle integrated electron distribution, $\int_0^{\pi}\sin\alpha_{eq}\cdot psd(E,\alpha_{eq})d\alpha_{eq}$ in a long-term simulation.}
\label{fig13}
\end{figure}

Moreover, note that our simulations shown in Figures \ref{fig9}-\ref{fig12} are quite localized in time, since $R/c\sim 1000$ is about 100 s in the outer radiation belt ($L\sim 5$), and that this time period is much smaller than the characteristic time of evolution of any process typically modelled by quasi-linear theory \citep{Thorne13:nature, Drozdov15, Albert16, Ma16:diffusion, Ma18}. Therefore, we extend the simulation interval to $tc/R=2.5\cdot10^4$ ($\sim 40$ min) to show that this time scale is already sufficiently long to almost fully smooth gradients within $<1$ MeV, see Fig. \ref{fig13}. Generally, however, 40 minutes is a too long interval to keep whistler-mode wave activity at the same high level (although such long-living regions of intense waves are sometimes observed, see \citep{Cully08, Agapitov15:grl:acceleration, Cattell15}).

Figure \ref{fig8} shows energy/pitch-angle domains of nonlinear wave-particle interaction, and these domains are used for the simulation of the electron distribution function evolution (see Figures \ref{fig10}-\ref{fig13} and Appendix C). For simplicity, we assume that the boundary of these domains is impenetrable. However, additionally to nonlinear wave-particle interactions (phase bunching and phase trapping), there is also in reality some electron diffusion. This diffusion is finite everywhere in the energy/pitch-angle plane where there is an electron resonance with the whistler-mode wave. Thus, this diffusion would transport electrons across the boundary of the domains of nonlinear wave-particle interaction. The direction of this transport depends on the phase space density gradients. At low energies, the nonlinear wave-particle interaction results in phase space density decrease (see Fig. \ref{fig13}), and thus pitch-angle diffusion will bring new small-energy particles into these domains. At high energies, the nonlinear wave-particle interaction results in phase space density increase (see Fig. \ref{fig13}), and thus both energy and pitch-angle diffusion will try to spread this phase space density maximum. Such diffusion can be included into the map (\ref{eq21}) as random energy and pitch-angle jumps with zero mean values and amplitudes given by the quasi-linear model \citep[e.g.,][]{Albert10}. However, the diffusion is generally much weaker than nonlinear phase bunching and trapping, and the diffusion-driven evolution of the phase space density should mostly appear after nonlinear wave-particle interaction has already partly smoothed the initial phase space density gradients \citep{Artemyev19:pd}.

The map (\ref{eq21}) has been constructed for electron interaction with monochromatic waves (see Eq. (\ref{eq01})), whereas spacecraft observations in the Earth’s radiation belts often report about more complex wave field distributions, e.g., significant wave amplitude modulation \citep{Tao13, Santolik14:rbsp, Zhang18:jgr:intensewaves, Zhang19:grl}, accompanied by fast, strong, and random variations of wave frequency and phase \citep{Zhang20:grl:frequency, Zhang20:grl:phase}, often resulting in the formation of almost independent short wave packets or sub-packets \citep{Mourenas18:jgr, Zhang20:grl:phase}. Such a chaotization of wave fields is likely partly driven by currents of resonant electrons \citep{Nunn09, Demekhov11, Katoh&Omura11, Katoh&Omura16, Tao17:generation, Tao20} and sideband instability \citep{Nunn86}, as well as by the simultaneous excitation of at least two different waves with a significant frequency difference \citep{Katoh&Omura13, Crabtree17:jgr, Zhang20:grl:frequency}. Since phase bunching is a local process, wave modulation cannot affect the theoretical model of energy and pitch-angle jumps due to bunching, but the inclusion of such a modulation into the 2-wave model map would require some probabilistic distribution of wave amplitudes within short wave packets. The situation is more complicated for phase trapping, which is nonlocal and depends on wave packet size and amplitude modulation within the packets \citep{Mourenas18:jgr}. Test particle simulations demonstrate that wave modulation alone makes phase trapping less efficient for electron acceleration, but increases the probability of phase trapping \citep{Kubota&Omura18, Gan20:grl, Zhang20:grl:phase}. Thus, an important further development of the mapping technique for nonlinear wave-particle interaction would require modifications of the phase trapping model.

To conclude, we have demonstrated the usefulness of the mapping technique for Hamiltonian systems describing nonlinear resonant interaction of charged particles and intense electromagnetic waves. We have shown that in  systems with two (and more) waves, the resonant interaction destroys the symmetries of the single wave resonance and drives a rapid smoothing of particle phase space density gradients. The proposed approach appears very promising for the investigation of relativistic electron interaction with various intense whistler-mode waves and EMIC waves in the Earth’s radiation belts \citep{Katoh&Omura13, Mourenas16, Mourenas16:grl, Ma17:vlf, He20, Yu20, Zhang20:grl:frequency} or in the solar wind \citep{Wilson07, Wilson13:waves, Krafft13, Krafft&Volokitin16, Tong19:ApJ, RobergClark19}. It could be useful also for studying electron acceleration by simultaneous laser-driven plasma waves \citep{Modena95, Tikhonchuk19}, and electron precipitation driven by VLF waves generated by electron beams or antennas in space \citep{Carlsten19, Borovsky20:aurora}. 

\section*{Acknowledgements}
The work of A.V.A., A.I.N., and A.A.V. was supported by Russian Scientific Foundation (project no. 19-12-00313). The work of A.V.A and X.J.Z. was supported in part by NSF grant 2021749 and NASA grant 80NSSC20K1270. The work of X.J.Z. and D.L.V. was supported by NASA grant 80NSSC20K1578. The work of A.I.N. was supported in part by the Leverhulme Trust grant RPG-2018-143.

\section*{Appendix A}
Equation (\ref{eq17}) describes energy decrease due to phase bunching, and natural limitation of this equation is that $\gamma+\Delta\gamma$ should be larger than one; or, alternatively, $I_x+(m_ec^2/\omega) \Delta \gamma$ should be larger than zero. This effect of drift asymmetry, i.e. absence of electron drift to negative $I_x$, has been noticed by \citet{Lundin&Shkliar77} who showed that for very small $I_x$ the phase bunching change the drift direction. This effect is called anomalous phase bunching \citep{Kitahara&Katoh19,Grach&Demekhov20, Gan20:grl} and basically consists in positive $I_x$ (and $\gamma$) changes due to bunching at very small $I_x$. Theoretically, the parametrical boundary of anomalous bunching in energy/pitch-angle space is determined by $I_x<I_x^*$ with $I_x^*$ scaling as $(B_w/B_0)^{2/3}$. Let us derive this scaling, but leave the more detailed consideration of small $I_x$ phase bunching to further consideration. We start with Eq. (\ref{eq05}) written for a single wave
\begin{equation}
 H_I  =  - \omega I  + m_e c^2 \gamma  + \sqrt {\frac{{2I\Omega _{ce} }}{{m_e c^2 }}} \frac{e}{\gamma }\frac{{B_{w} }}{{k}}\sin \varphi 
\label{eqA1},\;\;\;\;
 \gamma  = \sqrt {1 + \frac{{\left( {P + kI  } \right)^2 }}{{m_e^2 c^2 }} + \frac{{2I\Omega _{ce} }}{{m_e c^2 }}} 
 \end{equation}
Hamiltonian equations for $I$ and $\varphi$ take the form:
\begin{equation}
\dot I =  - \sqrt {\frac{{2I\Omega _{ce} }}{{m_e c^2 }}} \frac{{eB_w }}{{k\gamma }}\sin \varphi ,\quad \dot \varphi  = \frac{{k^2 }}{{\gamma m_e }}\left( {I - I_R } \right) + \sqrt {\frac{{\Omega _{ce} }}{{2Im_e c^2 }}} \frac{{eB_w }}{{k\gamma }}\sin \varphi 
\label{eqA2}
\end{equation}
where $I_R=(\gamma\omega m_e-P)/k$ is the solution of $\partial H/\partial I=0$ equation for $B_w=0$. Equation (\ref{eqA2}) describes fast phase rotation (with frequency $k^2(I-I_R)/\gamma m_e$) and $I,\varphi$ evolution driven by much weaker wave force $\sim B_w/B_0$. Until $I$ (and $I_R$) are sufficiently large to keep this time separation, we can apply the theory of phase bunching resulting in Eq. (\ref{eq17}). However, let us consider small $I, I_R$ values. We introduce a small parameter $\varepsilon=B_w/B_0$ and normalized $(\tilde{I},\tilde{I}_R)=(I,I_R)/\varepsilon^\beta$:
\begin{equation}
\frac{{d\tilde I}}{{dt}} =  - \sqrt {\frac{{2\tilde I\Omega _{ce} }}{{m_e c^2 }}} \frac{{eB_0 }}{{k\gamma }}\varepsilon ^{1 - \beta /2} \sin \varphi ,\quad \frac{{d\varphi }}{{dt}} = \frac{{k^2 \varepsilon ^\beta}}{{\gamma m_e }}\left( {\tilde I - \tilde I_R } \right)   + \sqrt {\frac{{\Omega _{ce} }}{{2\tilde Im_e c^2 }}} \frac{{eB_0 }}{{k\gamma }}\varepsilon ^{1 - \beta /2} \sin \varphi 
\label{eqA3}
\end{equation}
Introducing slow time $\tau=t\varepsilon^{1-\beta/2}$, we obtain	
\begin{equation}
\frac{{d\tilde I}}{{d\tau }} =  - \sqrt {\frac{{2\tilde I\Omega _{ce} }}{{m_e c^2 }}} \frac{{eB_0 }}{{k\gamma }}\sin \varphi ,\quad \frac{{d\varphi }}{{d\tau }} = \frac{{k^2 }}{{\gamma m_e }}\left( {\tilde I - \tilde I_R } \right)\varepsilon ^{3\beta /2 - 1}  + \sqrt {\frac{{\Omega _{ce} }}{{2\tilde Im_e c^2 }}} \frac{{eB_0 }}{{k\gamma }}\sin \varphi 
\label{eqA4}
\end{equation}
Thus, for $\beta=2/3$ Eqs. (\ref{eqA4}) lose the small parameter, and $\tilde{I},\varphi$ would change with the same rate. Then the applicability of equations of the phase bunching theory breaks, and a new model for $\Delta I$ (or $\Delta\gamma$, $\Delta I_x$) is required. $\beta=2/3$ gives the threshold value for $I_x\sim I\sim (B_w/B_0)^{\beta}$.

\section*{Appendix B}
\begin{figure}
\centerline{\includegraphics[width=1.0\textwidth]{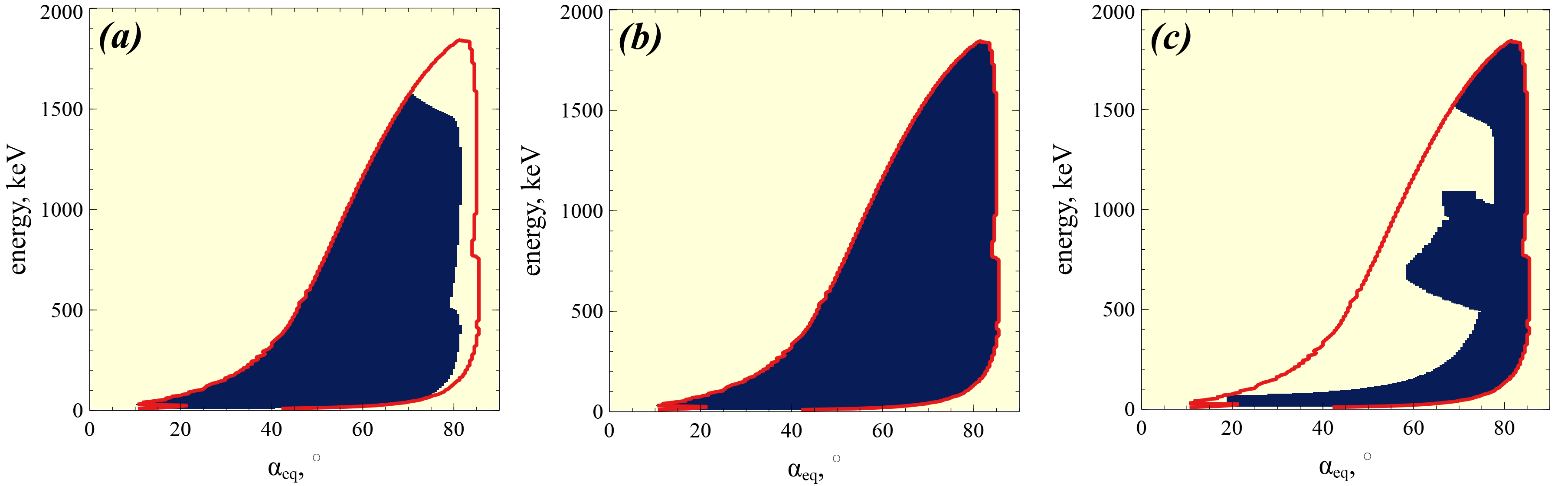}}
  \caption{Domains in energy/pitch-angle space with: positive probability of trapping (a), a finite phase bunching energy/pitch-angle change (b), positions of  release from the trapping (c). Red curve shows boundary of (b) domain.}
\label{fig14}
\end{figure}

Figure \ref{fig8} shows that there are certain domains in the energy/pitch-angle space where electrons resonate with whistler wave nonlinearly. Thus, simulation of resonant electron dynamics should be within these domains. Figure \ref{fig14}(b) shows the largest domain that cover all energies and pitch-angles where electrons experience phase bunching. The phase bunching results in energy/pitch-angle change and electron drifts within the domain. Important property of the domain boundary is that $\Delta_{scat}\gamma$ tends there to zero as $\sim (I-I_{boundary})^{4/3}$ where $I$ and $I_{boundary}$ are values of moment and it’s boundary value \citep{Artemyev19:pd}, i.e. $\Delta\gamma$ drops to zero at the domain boundary and no particles should leave this domain (in absence of diffusion that is characterized by a finite diffusion coefficient within the entire energy/pitch-angle space). As $\Delta_{scat}\gamma$ has been derived numerically, there are possible fluctuations making $\Delta_{scat}\gamma$ finite at the boundary. Thus, distribution $\Delta\gamma (E,\alpha_{eq})$ should be corrected to set $\Delta_{scat}\gamma=0$ at the domain boundary. Moreover, if during the simulation resonant electrons escape from the domain of phase bunching (e.g., because of numerical effects), these electrons should be returned into the domain (e.g., reflecting them back from the boundary on the same $\Delta_{scat}\gamma$). Note that this procedure is required only in the absence of particle diffusion.
\footnote{The system has 3 degrees of freedom. Thus, generally there are deviations from  adiabatic trajectories even for non-resonant motions due to the Arnold diffusion (\cite{bookAKN06}). However, this diffusion is exponentially slow and can be neglected here.}

The domain of a finite trapping probability is smaller than the bunching domain (see Fig. \ref{fig14}(a)). Again, the probability of trapping tends to zero at the phase bunching domain boundary as $\Pi\sim (I-I_{boundary})^{1/3}$ \citep{Artemyev19:pd}, and $\Pi$ should be set equal to zero on this boundary even if numerical fluctuations of $\Pi$ evaluation give some finite value. Of course, there are no regions with $\Pi>0$ outside the phase bunching domain. 

Release of trapped electrons from the resonance also should be within the phase bunching domain (see Fig. \ref{fig14}(c)). Numerical errors put some release locations outside this domain; the trapping variation $\Delta_{trap}\gamma$ should be corrected to move the release locations within the domain. This guarantees that for each energy/pitch-angle within the phase bunching domain we would have incoming and outcoming phase space flows.

\end{document}